\documentclass[aps,prb,amsmath,amssymb,reprint,superscriptaddress,preprintnumbers,showpacs,intlimits,longbibliography]{revtex4-1}
\pdfoutput=1
\bibpunct{[}{]}{,}{n}{}{}

\usepackage[utf8]{inputenc}
\usepackage{titlesec}
\usepackage{bm,latexsym,mathrsfs,enumerate}
\usepackage[dvipsnames]{xcolor}
\usepackage{mathtools}
\usepackage{makecell}
\usepackage{multirow}
\usepackage[breaklinks=true,unicode=true,urlcolor = blue,colorlinks = true,citecolor = blue,linkcolor = blue]{hyperref}
\usepackage{graphicx}
\usepackage{ragged2e}
\renewcommand{\vec}[1]{\bm{#1}}
\begin{document}

\title{Spontaneous deformation of flexible ferromagnetic ribbons induced by Dzyaloshinskii--Moriya interaction}

\author{Kostiantyn~V.~Yershov}
\email{yershov@bitp.kiev.ua}
\affiliation{Bogolyubov Institute for Theoretical Physics of National Academy of Sciences of Ukraine, 03143 Kyiv, Ukraine}
\affiliation{Leibniz-Institut f\"ur Festk\"orper- und Werkstoffforschung, IFW Dresden, D-01171 Dresden, Germany}

\author{Volodymyr~P.~Kravchuk}
\email{vkravchuk@bitp.kiev.ua}
\affiliation{Bogolyubov Institute for Theoretical Physics of National Academy of Sciences of Ukraine, 03143 Kyiv, Ukraine}
\affiliation{Leibniz-Institut f\"ur Festk\"orper- und Werkstoffforschung, IFW Dresden, D-01171 Dresden, Germany}
\affiliation{Institut f\"ur Theoretische Festk\"orperphysik, Karlsruher Institut f\"ur Technologie, D-76131 Karlsruhe, Germany}

\author{Denis~D.~Sheka}
\email{sheka@knu.ua}
\affiliation{Taras Shevchenko National University of Kyiv, 01601 Kyiv, Ukraine}

\author{Jeroen van den Brink}
\email{j.van.den.brink@ifw-dresden.de}
\affiliation{Leibniz-Institut f\"ur Festk\"orper- und Werkstoffforschung, IFW Dresden, D-01171 Dresden, Germany}
\affiliation{Institute for Theoretical Physics, TU Dresden, 01069 Dresden, Germany}
\affiliation{Department of Physics, Washington University, St. Louis, MO 63130, USA}

\author{Yuri Gaididei}
\email{ybg@bitp.kiev.ua}
\affiliation{Bogolyubov Institute for Theoretical Physics of National Academy of Sciences of Ukraine, 03143 Kyiv, Ukraine}

\begin{abstract}

Here, we predict the effect of the spontaneous deformation of a flexible ferromagnetic ribbon induced by Dzyaloshinskii--Moriya interaction~(DMI). The geometrical form of the deformation is determined both by the type of DMI and by the equilibrium magnetization of the stripe. We found three different geometrical phases, namely (i) the DNA-like deformation with the stripe central line in the form of a helix, (ii) the helicoid deformation with the straight central line and (iii) cylindrical deformation. In the main approximation the magnitude of the DMI-induced deformation is determined by the ratio of the DMI constant and the Young's modulus. It can be effectively controlled by the external magnetic field, what can be utilized for the nanorobotics applications. All analytical calculations are confirmed by numerical simulations.

\end{abstract}
\maketitle


\section{Introduction}

Magnetic soft matter opens new possibilities in construction and fabrication of shapeable magnetoelectronics~\cite{Makarov16,Sheng18}, interactive human-machine interfaces~\cite{Wang18,Hu18}, and programmable magnetic materials~\cite{Lum16,Kim18a}. Remote control of the shape and 3D navigation of the soft magnet by means of the external magnetic field stimulate intensive investigations in the area of milli-~\cite{Hu18,Lum16,Kim18a,Lu18,Zhang18d} and microrobotics~\cite{Kim11,Medina-Sanchez18,Ceylan19} for flexible electronics and biomedical applications. So far the  magneto-sensitive elastomers~\cite{Singh05,Thevenot13,Herzer13,Geryak14,Townsend14,Lopatina15} are the most studied magnetically responsive flexible materials. The magnetic properties of elastomers are determined by the long range dipole-dipole interaction~\cite{Romeis17,VazquezMontejo18,Brisbois19,Zhao19a} which results in the relatively large scale of the geometrical deformations. It is well known that organic, organic-inorganic hybrid, and molecule-based magnets exhibiting different types of magnetic ordering  \cite{Ovchinnikov88,Miller88,Palacio93,Miller02,Podgajny06,Barron08,Miller11,Miller14} and some of them can keep ferromagnetic order even for a room temperature~\cite{Mahmood18}. In comparison with elastomers, the dominant interaction in the molecule-based magnets is a local short-range exchange interaction. Therefore, the scale of deformations in such systems is in nanoscale range which allows significantly reduce the size of the object. The sub-micrometer size of the molecule-based magnets and possibility to control the geometry of the magnet by means of magnetic field opens new possibilities in the development of \textit{nanorobots} in the context of organic electronics and spintronics \cite{Bujak13}.

Deformation of a flexible magnet induced by its magnetization subsystem was predicted in a number of previous works~\cite{Dandoloff95,Saxena97,Saxena98,Gaididei19}. Here, we demonstrate that a presence of intrinsic DMI results in a spontaneous deformation of a flexible magnetic ribbon. Depending on the mechanical, magnetic, geometric parameters, and the symmetry of the DMI one can obtain different equilibrium states, see Table~\ref{tab:states}.  A promising feature of the DMI induced deformation is its field-controlled reconfigurability, which is an important issue for the nanorobotics applications. The numerical simulations confirm our analytical calculations: shape of the deformed ribbon, phase diagram of equilibrium states. We used an in-house developed simulating code, which takes into account both magnetic and geometrical degrees of freedom.


\begin{table}[b]
	\begin{tabular}{p{0.05\columnwidth}|p{0.04\columnwidth}|p{0.4\columnwidth}|p{0.4\columnwidth}|}
		\multicolumn{2}{c|}{} & \multicolumn{2}{c|}{DMI type} \\
		\hline
		\multirow{5}{*}[-1.3cm]{\rotatebox{90}{Magnetization direction}} &  \multirow{2}{*}[-1.25cm]{\rotatebox{90}{$\vec{m}=\pm\vec{e}_1$}} & \centering $\mathcal{E}_\textsc{d}^\textsc{b}=\vec{m}\cdot\left[\vec{\nabla}\times\vec{m}\right]$ & $\mathcal{E}_\textsc{d}^\textsc{n}=m_n \vec{\nabla}\cdot\vec{m}-\vec{m}\cdot\vec{\nabla}m_n$  \\\cline{2-4}
		& & \centering DNA-like and helicoid states & $\quad$ Undeformed state \\ 
		&  & \centering \includegraphics[width=0.25\columnwidth]{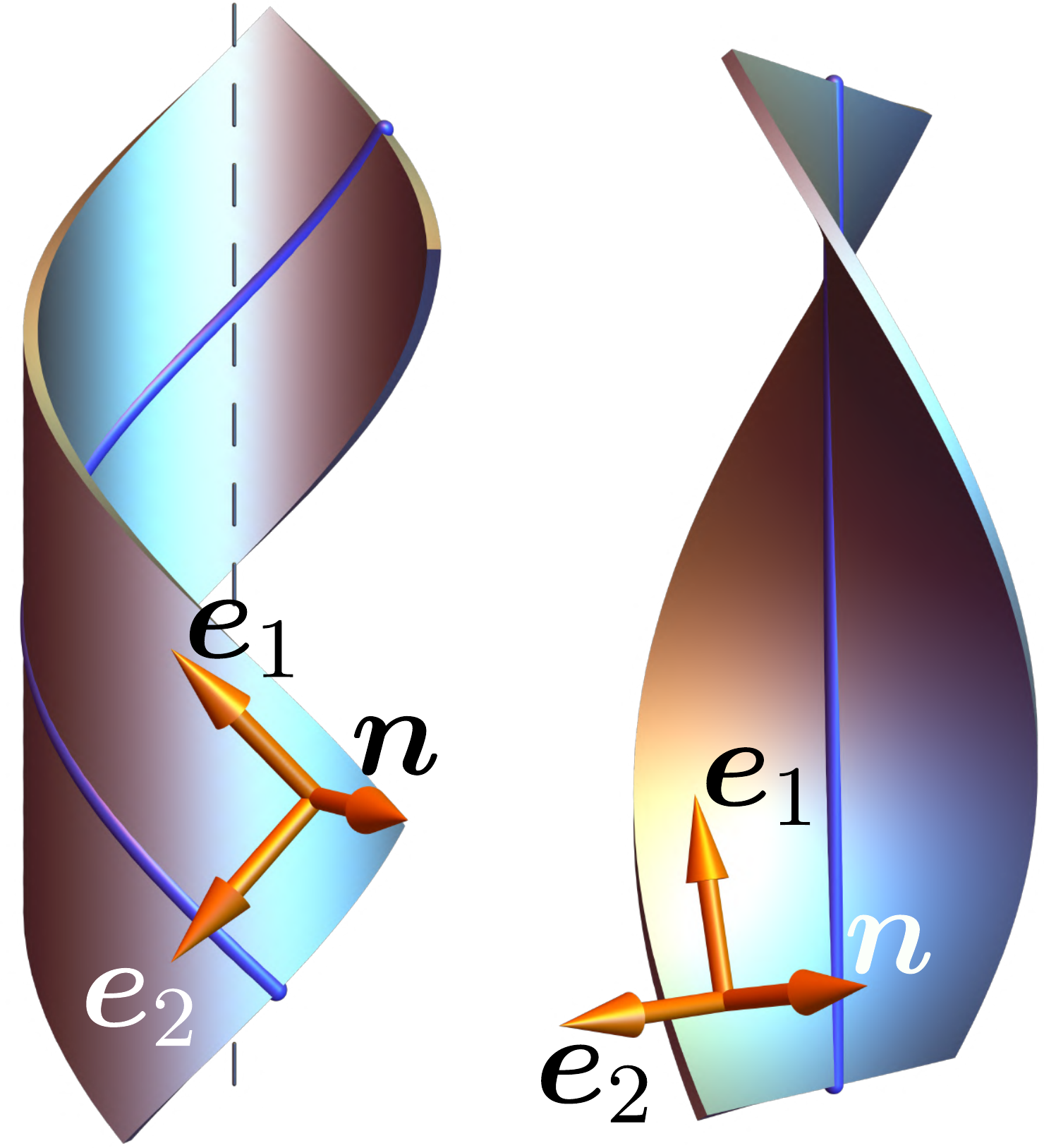} & \includegraphics[width=0.4\columnwidth]{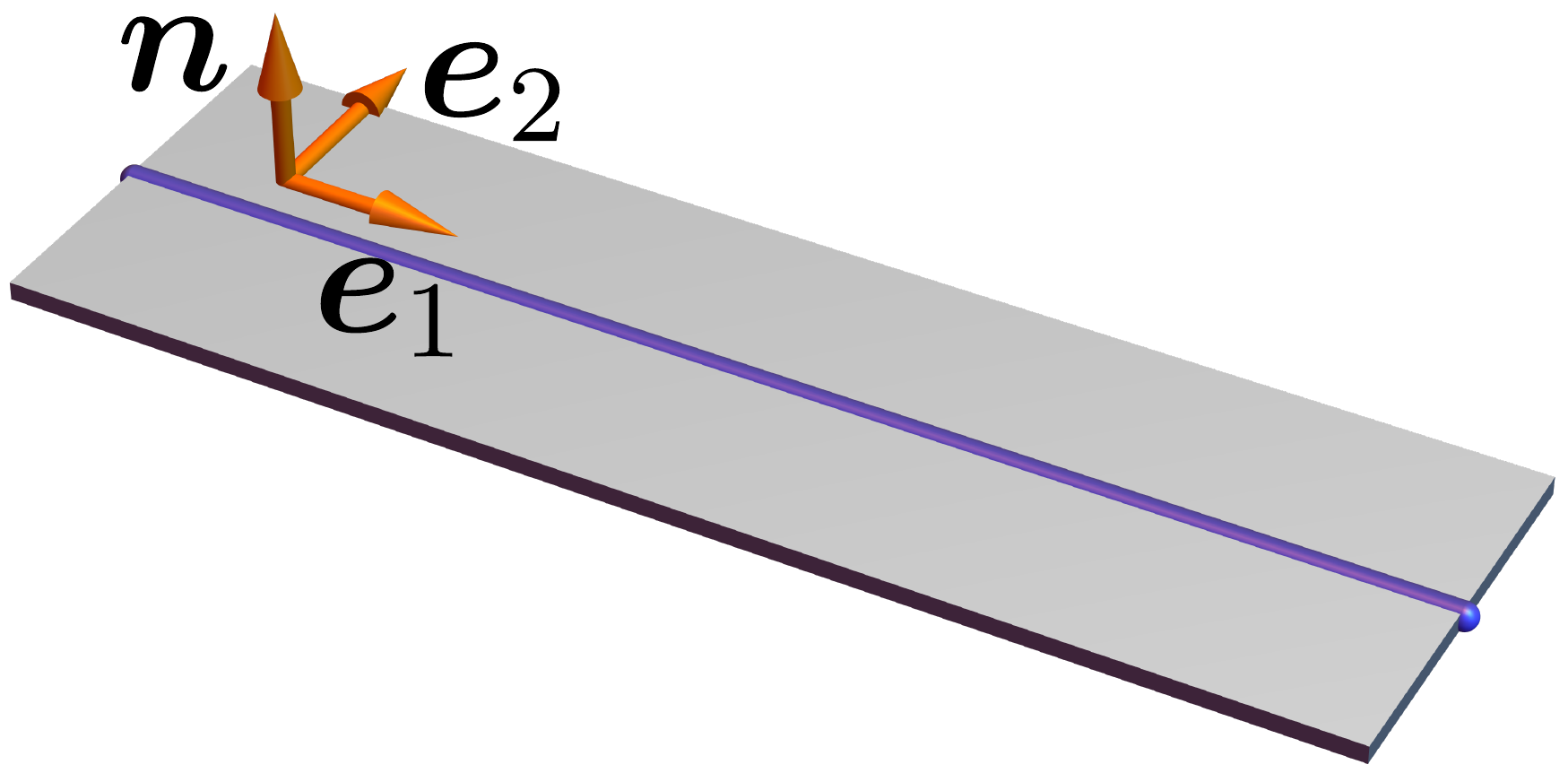} \\\cline{2-4} 
		& \multirow{2}{*}[-0.5cm]{\rotatebox{90}{$\vec{m}=\pm\vec{n}$}} & $\quad$ Undeformed state & $\quad$ Cylindrical state\\
		&  & \includegraphics[width=0.4\columnwidth]{fig_flat.pdf} & \includegraphics[width=0.4\columnwidth]{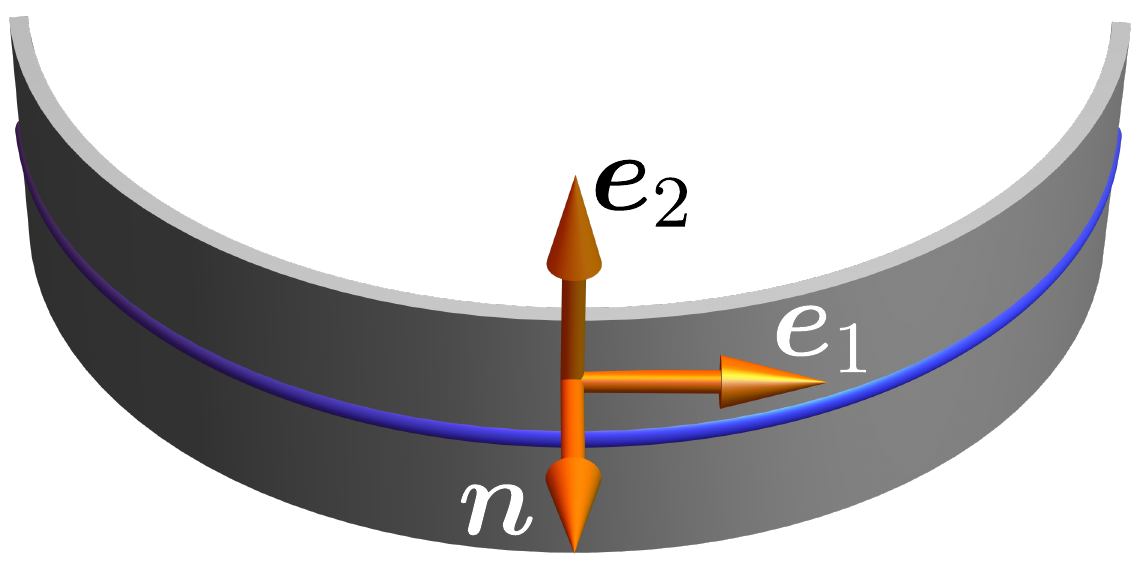}\\
		\hline 
	\end{tabular} 
	\caption{\label{tab:states}Schematic illustrations of possible DMI-induced deformations of flexible ferromagnetic ribbon for different types of DMI and directions of the magnetization.}
\end{table}

\begin{figure*}
	\includegraphics[width=\textwidth]{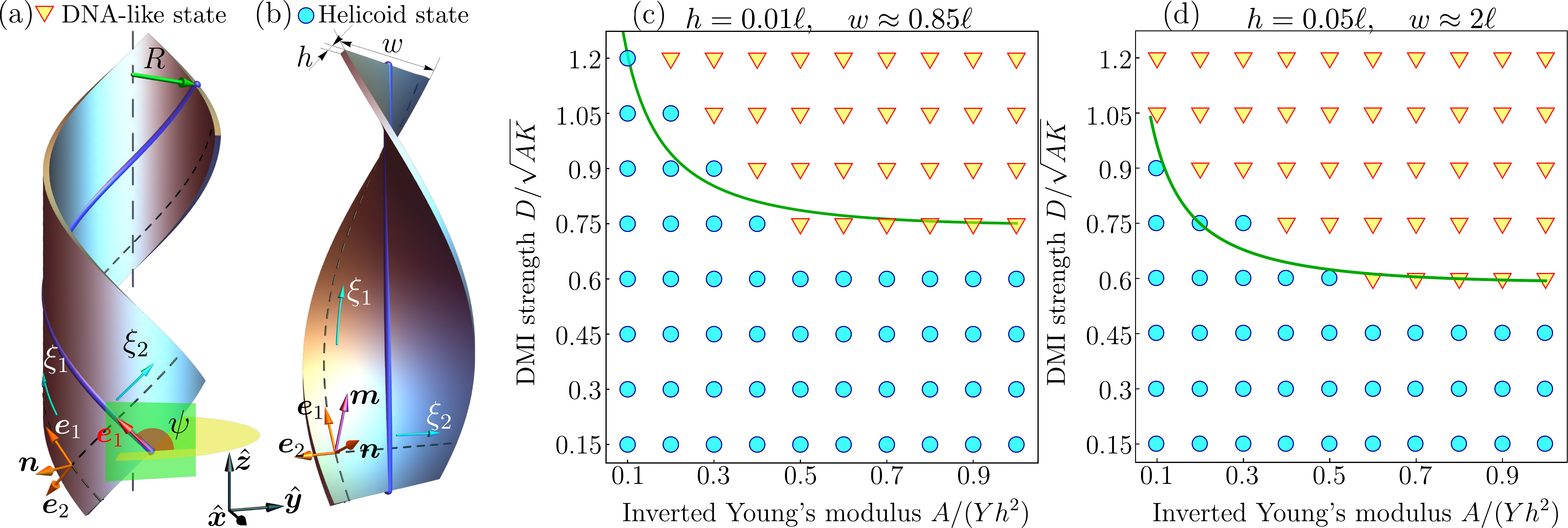}
	\caption{\label{fig:states}%
		(Color online) \textbf{Equilibrium states of flexible ferromagnetic ribbon with DMI in form} $\mathcal{E}_\textsc{d}=\mathcal{E}_\textsc{d}^\textsc{b}$\textbf{:} (a,b) Shapes of DNA-like~(a) and helicoid~(b) states. (c) and (d) are phase diagrams of equilibrium states of the flexible ribbon. Symbols show the results of the numerical simulations: circles and triangles correspond to helicoid and DNA-like states, respectively. Thick green line in (c) and (d) describes the boundary between equilibrium states, see Appendix~\ref{sec:bulk_dmi} for details. In all cases we have~$\nu=1/3$.}
\end{figure*} 

\section{Model}

We consider a 3D narrow ferromagnetic ribbon of rectangular cross section  whose thickness $h$ and width $w$ are small enough to ensure the magnetization uniformity along a ribbon cross section. The ribbon length $L$ is substantially larger than the transversal dimensions~($h\ll w\ll L$).  The space domain occupied by the ribbon is defined as $\vec{r}(\xi_1,\xi_2,\eta) = \vec{\varsigma}(\xi_1,\xi_2) + \eta\, \vec{n}(\xi_1,\xi_2)$. Here, $\vec{\varsigma}$ determines a 2D surface $\mathcal{S}$ embedded in $\mathbb{R}^3$ with  $\xi_1\in[0,L]$ and $\xi_2\in[-w/2,w/2]$ being local curvilinear coordinates on $\mathcal{S}$. The unit vector $\vec{n}$ denotes the surface normal and the parameter $\eta\in[-h/2,h/2]$ is the curvilinear coordinate along the normal direction. The parametrization $\vec{\varsigma}(\xi_1,\xi_2)$ induces the natural tangential basis $\vec{g}_\alpha = \partial_\alpha\vec{\varsigma}$ with the corresponding metric tensor elements $g_{\alpha\beta}=\vec{g}_\alpha\cdot\vec{g}_\beta$. Here, $\alpha,\beta=1,2$ and $\partial_\alpha\equiv\partial_{\xi_\alpha}$. Assuming that vectors $\vec{g}_\alpha$ are orthogonal, one can introduce the orthonormal basis $\{\vec{e}_1, \vec{e}_2, \vec{n}\}$, where $\vec{e}_\alpha=\vec{g}_\alpha/\sqrt{g_{\alpha\alpha}}$ and $\vec{n}=\vec{e}_1\times\vec{e}_2$, see Fig.~\ref{fig:states}(a)-(b) for detailed notations.

\begin{figure*}
	\includegraphics[width=\textwidth]{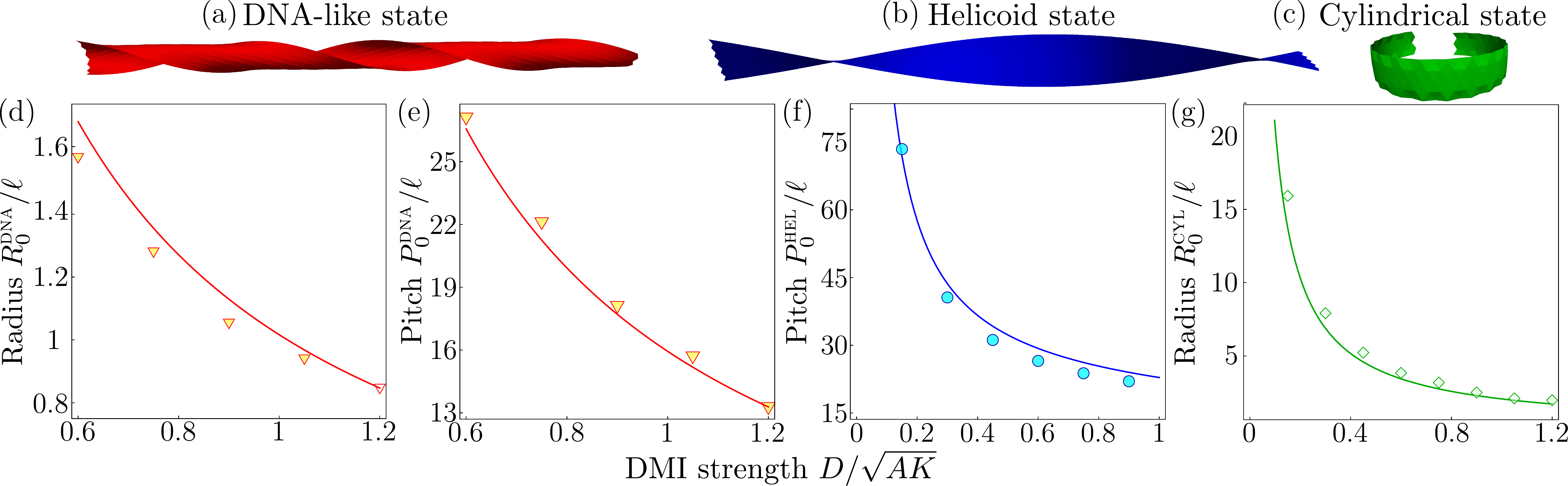}
	\caption{\label{fig:parameters}%
		(Color online) \textbf{Shapes and geometrical parameters of flexible ferromagnetic ribbons as functions of DMI strength}: (a) -- (c) are the ribbon shapes obtained by means of numerical simulations. (d),(e) are radius and pitch of DNA-like state plotted for $A/\left(Yh^2\right)=1$, respectively. (f) pitch of the helicoid state for $A/\left(Yh^2\right)=0.11$. (g) Radius of cylindrical state for $A/\left(Yh^2\right)=1$. Red lines in (d),(e) and green line in (g) represent analytical predictions~\eqref{eq:dna_equilibrium} and~\eqref{eq:radius_cylinder}, respectively. Blue line in (f) is obtained from solution of the cubic Eq.\eqref{eq:hel_param_def_suppl}. Symbols correspond to data obtained by means of numerical simulations. All data is presented for ribbons with width $w\approx2\ell$, thickness $h=0.05\ell$, and Poisson ratio $\nu=1/3$. The dynamical process of the DMI induced deformation is illustrated in the Supplemental movie~\ref{sec:video_bloch_free}.}
\end{figure*} 

The total energy $E=E_\textsc{e}+E_\textsc{m}$ of a flexible ferromagnetic ribbon is a summation of elastic~\cite{Efrati09,Armon11}
\begin{subequations} \label{eq:model}
\begin{equation}\label{eq:model_elastic}
E_\textsc{e}=\frac{Y}{8(1+\nu)}\int\limits_{0}^{L}\int\limits_{-w/2}^{w/2}\left(h\mathcal{E}_\textsc{s}+\frac{h^3}{3}\mathcal{E}_\textsc{b}\right)\sqrt{\overline{g}}\textrm{d}\xi_1\mathrm{d}\xi_2
\end{equation}
and magnetic 
\begin{equation}\label{eq:model_magnetic}
E_\textsc{m}=h\int\limits_{0}^{L}\int\limits_{-w/2}^{w/2}\left(A\mathcal{E}_\textsc{ex}+K\mathcal{E}_\textsc{a}+D\mathcal{E}_\textsc{d}\right)\sqrt{g}\textrm{d}\xi_1\mathrm{d}\xi_2
\end{equation}
energy terms. Elastic energy is taken for the case of thin amorphous films where only terms of first and third order of magnitude with respect to thickness $h$ are taken into account~\cite{Efrati09}. Here, $g=\det\|g_{\alpha\beta}\|$ and  $\overline{g}=\det\|\overline{g}_{\alpha\beta}\|$	with $\overline{g}_{\alpha\beta}$ being the metric tensor for ribbon free of elastic tensions (we consider a straight ribbon with $\overline{g}_{\alpha\beta}=\delta_{\alpha\beta}$ as a reference metric). Parameters $Y$ and $\nu\in\left[0,0.5\right]$ in \eqref{eq:model_elastic} are Young's modulus and Poisson ratio, respectively.
\end{subequations}

The first term in elastic energy~\eqref{eq:model_elastic} determines stretching energy density $\mathcal{E}_\textsc{s}=\left(\frac{\nu}{1-\nu}\overline{g}^{\alpha\beta}\overline{g}^{\gamma\delta}+\overline{g}^{\alpha\gamma}\overline{g}^{\beta\delta}\right)\left(g_{\alpha\beta}-\overline{g}_{\alpha\beta}\right)\left(g_{\gamma\delta}-\overline{g}_{\gamma\delta}\right)$. The last term in \eqref{eq:model_elastic} corresponds to the bending energy $\mathcal{E}_\textsc{b}=\left(\frac{\nu}{1-\nu}\overline{g}^{\alpha\beta}\overline{g}^{\gamma\delta}+\overline{g}^{\alpha\gamma}\overline{g}^{\beta\delta}\right)b_{\alpha\beta}b_{\gamma\delta}$ with $b_{\alpha\beta}=\vec{n}\cdot\partial_\beta\vec{g}_\alpha$ being the second fundamental form.

The first term in \eqref{eq:model_magnetic} is the exchange energy density with $\mathcal{E}_\textsc{ex}=\sum_{i=x,y,z}\left(\partial_i\vec{m}\right)^2$, and $A$ is an exchange constant. Here $\vec{m}=\vec{M}/M_s$ is the unit magnetization vector with $M_s$ being the saturation magnetization. The second term in \eqref{eq:model_magnetic} is the anisotropy energy density $\mathcal{E}_\textsc{a} = 1 - \left(\vec{m}\cdot\vec{e}_\textsc{a}\right)^2$ with $\vec{e}_\textsc{a}$ being easy-axis vector. The vector $\vec{e}_\textsc{a}$ follows either normal or tangential direction and in this way, the anisotropy term in~\eqref{eq:model_magnetic} realizes the magneto-elastic coupling. Parameter $K>0$ is easy-axial anisotropy constant. The exchange-anisotropy competition results in the magnetic length $\ell=\sqrt{A/K}$, which determines length scale of the system. The last term in~\eqref{eq:model_magnetic} represents DMI contribution  $\mathcal{E}_\textsc{d}$ with $D$ being the DMI constant. We consider two types of DMI: (i) $\mathcal{E}_\textsc{d}^\textsc{b}=\vec{m}\cdot\left[\vec{\nabla}\times\vec{m}\right]$ is typical for systems with $T$ symmetry~\cite{Cortes-Ortuno13}. In the following we call this DMI of Bloch type, since it results in the domain walls and skyrmions of Bloch type. (ii) $\mathcal{E}_\textsc{d}^\textsc{n}=m_n \vec{\nabla}\cdot\vec{m}-\vec{m}\cdot\vec{\nabla}m_n$ is typical for ultrathin films~\cite{Bogdanov01,Thiaville12},
bilayers~\cite{Yang15} or materials belonging to $C_{nv}$ crystallographic group. In the following we call this DMI of N{\'e}el type. Recently it was shown that Ne{\'e}l DMI can be obtained in the Janus monolayers of chromium trihalides Cr(I,X)$_3$~\cite{Xu19}. It is also important to note that DMI was recently observed in amorphous GdFeCo films~\cite{Kim19}. 

A DMI in a rigid magnetic system results in the appearance of periodical structures~(e.g. conical or helical modulations~\cite{Dzyaloshinski64,Dzyaloshinski65,Bogdanov05,Heide11}). In systems with strong enough anisotropy~($|D|/\sqrt{AK}<4/\pi$) the periodical structures are suppressed and we have uniform magnetization distribution. However, if we add additional elastic degree of freedom to the system, one should expect realization of a periodical magnetization distribution due to the 3D deformation of the ribbon. Schematic illustrations of the found DMI-induced deformations of flexible ferromagnetic ribbons are presented in Table~\ref{tab:states}.

Now we utilize the model \eqref{eq:model} to provide an analytic description for the DMI induced flexible ferromagnetic ribbon deformation.

\section{DMI of Bloch type}

Here we consider DMI in form  $\mathcal{E}_\textsc{d}=\mathcal{E}_\textsc{d}^\textsc{b}$. We start with a tangential easy-axial anisotropy~($\vec{e}_\textsc{a}=\vec{e}_1$)~\footnote[2]{For ribbons with bulk DMI of Bloch type and normal easy-axial anisotropy, i.e. $\vec{e}_\textsc{a}=\vec{n}$, the equilibrium magnetization is aligned with the normal direction and DMI does not deforms the shape of the ribbon.}. Basing on our numerical simulations we assume that DMI induced deformation leads to the formation of two equilibrium states, namely: DNA-like~[Fig.~\ref{fig:parameters}(a)] and helicoid~[Fig.~\ref{fig:parameters}(b)] states. We start with a DNA-like state. Such a state is parameterized as $\vec{\varsigma}^\textsc{dna}\left(\xi_1,\xi_2\right)=R\cos\left(\rho/R\right)\hat{\vec{x}}+R\sin\left(\rho/R\right)\hat{\vec{y}}+\left(\xi_1\sin\psi+\xi_2\cos\psi\right)\hat{\vec{z}}$, where $\rho=\xi_1\cos\psi-\xi_2\sin\psi$, $R$ is a radius of the central line, and $\psi$ is an angle between vector $\vec{e}_1$ and $\vec{xy}$-plane, see Fig.~\ref{fig:states}(a). The pitch of the DNA-like state is $P=2\pi R\tan\psi$. And sign of the pitch determines the geometrical chirality $\mathcal{C}^\textsc{dna}=\text{sign}P^\textsc{dna}=\pm 1$. This parameterization results in the  Euclidean metrics, therefore this state is free of the stretching~(i.e. $g_{\alpha\beta}=\overline{g}_{\alpha\beta}$).

We show that the total energy~\eqref{eq:model} is minimized by a stationary solution $\vec{m}_0^\textsc{dna}=\mathfrak{C}\, \vec{e}_1$, where $\mathfrak{C}=\pm1$ determines whether magnetization is parallel~($\mathfrak{C}=1$) or antiparallel ($\mathfrak{C}=-1$) to the tangential axis (see App.~\ref{sec:bulk_dmi}). Equilibrium values for the radius $R$ and pitch $P$ are determined as
\begin{equation}\label{eq:dna_equilibrium}
R_0^\textsc{dna} = \dfrac{A}{|D|}\frac{2\sqrt[4]{1+\zeta}}{\sqrt{1+\zeta} - 1},\quad P_0^\textsc{dna} = \frac{A}{D} \frac{4\pi \sqrt{1+\zeta}}{\sqrt{1+\zeta}-1},
\end{equation}
with $\zeta=24\left(1-\nu^2\right)A/\left(Yh^2\right)$. Radius and pitch for the DNA-like state as functions of DMI strength are presented in~Figs.~\ref{fig:parameters}(d)-(e). For the case of relatively large values of Young's modulus ($A/\left(Yh^2\right)\ll1$) we have $R_0^\textsc{dna}\propto Yh^2/\left[6|D|\left(1-\nu^2\right)\right]$ and $P_0^\textsc{dna}\propto \pi Yh^2/\left[3D\left(1-\nu^2\right)\right]$.

The energy of the DNA-like state is
\begin{equation}\label{eq:dna_energy}
\frac{E_0^\textsc{dna}}{hwL}= -\frac{D^2}{4A}\frac{\sqrt{1+\zeta}-1}{\sqrt{1+\zeta}+1}\approx-3\left(1-\nu^2\right)\frac{D^2}{Yh^2}.
\end{equation}

The second equilibrium state is referred as a helicoid state. Such state can be parametrized in the following way $\vec{\varsigma}^\textsc{hel}\left(\xi_1,\xi_2\right)=\xi_2\left[\cos\left(k\xi_1\right)\hat{\vec{x}}+\sin\left(k\xi_1\right)\hat{\vec{y}}\right]+\xi_1\hat{\vec{z}}$, where $k$ is a twist parameter of the helicoid ribbon, which results in the pitch $P^\textsc{hel}=2\pi/k$. The helical state is also characterized by the geometrical chirality $\mathcal{C}^\textsc{hel}=\text{sign}P^\textsc{hel}=\pm 1$. The metric tensor for this state has a diagonal form $\|g_{\alpha\beta}\|=\text{diag}\left(1+k^2\xi_2^2,1\right)$. In contrast to the DNA-like state the helicoid geometry has nonzero Gau{\ss} curvature. This means that the metric tensor can not be transformed to the euclidean form. In our case it results in the stretching term in the energy~\eqref{eq:model}.

By minimizing enerrgy~\eqref{eq:model}, we obtained similar solution for the magnetization as for DNA-like state:  $\vec{m}_0^\textsc{hel}=\mathfrak{C}\, \vec{e}_1$. The equilibrium value of pitch for the case of narrow ribbons~$kw\ll1$ and large Young's modulus can be determined as
\begin{equation}\label{eq:twist_equilibrium}
P_0^\textsc{hel}\approx\frac{\pi}{3\left(1+\nu\right)}\frac{Yh^2}{D}\left[1+12\left(1+\nu\right)\frac{A}{Yh^2}\right].
\end{equation} 
Pitch of the helicoid state as a function of the DMI constant is presented in Figs.~\ref{fig:parameters}(f). 

The energy of the helicoid state is
\begin{equation}\label{eq:twist_energy}
\begin{split}
\frac{E_0^\textsc{hel}}{hwL}\approx-3\left(1+\nu\right)\frac{D^2 }{Yh^2}\biggl[1-\frac{27}{40}\frac{D^2}{Y^2h^2}\frac{w^4}{h^4}\frac{\left(1+\nu\right)^2}{\left(1-\nu\right)}\biggr].
\end{split}
\end{equation}

One should note that geometrical chirality of both states (DNA-like and helicoid) does not depend on the magnetization orientation and is defined only by the sign of DMI constant: $D>0$ for a left-handed ribbon and $D<0$ for a right-handed ribbon.

The helicoid state appears due to realization of conical phase allowed by the elastic degree of freedom. This state is characterized by nonzero stretching. The competition between the stretching and bending energies results in the appearance of the DNA-like state for the larger $D$. In the limit of small $D$ both energies $E_0^\textsc{hel}\propto-D^2$ and $E_0^\textsc{dna}\propto-D^2$ demonstrate quadratic dependence on $D$ and $E_0^\textsc{hel}<E_0^\textsc{dna}$. However, for larger $D$ the stretching induced term $\propto +D^4$ in \eqref{eq:twist_energy} results in the preferability of the DNA state $E_0^\textsc{dna}<E_0^\textsc{hel}$. By comparing energies of different states, we find the energetically preferable states for different $D$ and $Y$ values.  The resulting phase diagrams are presented in Fig.~\ref{fig:states}(c)-(d). There are two phases: (i) The DNA-like state is energetically preferable for relatively large values of $D$ or wide ribbons. (ii) The helicoid state is realized for relatively small values of $D$ or narrow ribbons. The magnetization distribution in both states is uniform in curvilinear reference frame and it is tangential to the ribbon surface. The boundary between two phases can be derived by using the condition $E_0^\textsc{hel}\left(D_c,Y\right)=E_0^\textsc{dna}\left(D_c,Y\right)$. The spontaneous deformations into the DNA-like and helicoid states are demonstrated in the supplemental movie~\ref{sec:video_bloch_free}.

\section{DMI of N{\'e}el type}

Here we consider DMI in form $\mathcal{E}_\textsc{d}=\mathcal{E}_\textsc{d}^\textsc{n}$ which is expected to Janus monolayers of Cr(I,Br)$_3$ and Cr(I,Cl)$_3$~\cite{Xu19}.  For ribbons with tangential easy-axial anisotropy, i.e. easy axis is oriented along the ribbon $\vec{e}_\textsc{a}=\vec{e}_1$, the equilibrium magnetization is aligned with the tangential direction and DMI does not deform the shape of the ribbon. While for the easy-normal anisotropy, DMI results in the deformation to the cylindrical structure, see Fig.~\ref{fig:parameters}(c). This deformation is a limit case of a DNA-like state with $\sin\psi_0^\textsc{dna}=0$. The equilibrium value of the radius is (see App.~\ref{sec:interfacial_dmi})
\begin{equation}\label{eq:radius_cylinder}
R_0^\textsc{cyl}=2\frac{A}{|D|}\left[1+\frac{Yh^2}{24A\left(1+\nu\right)}\right].
\end{equation}
Magnetization in this state is normal to the surface, i.e. $\vec{m}_0^\textsc{cyl}=\pm\vec{n}$. The energy of this state behaves as $E_0^\textsc{cyl}\propto -D^2/\left(Yh^2\right)$. The obtained prediction~\eqref{eq:radius_cylinder} is in a good agreement with numerical simulations, see Fig.~\ref{fig:parameters}(g). The spontaneous deformation into the cylindrical state is demonstrated in the supplemental movie~\ref{sec:video_neel_free}.

\begin{figure}
	\includegraphics[width=\columnwidth]{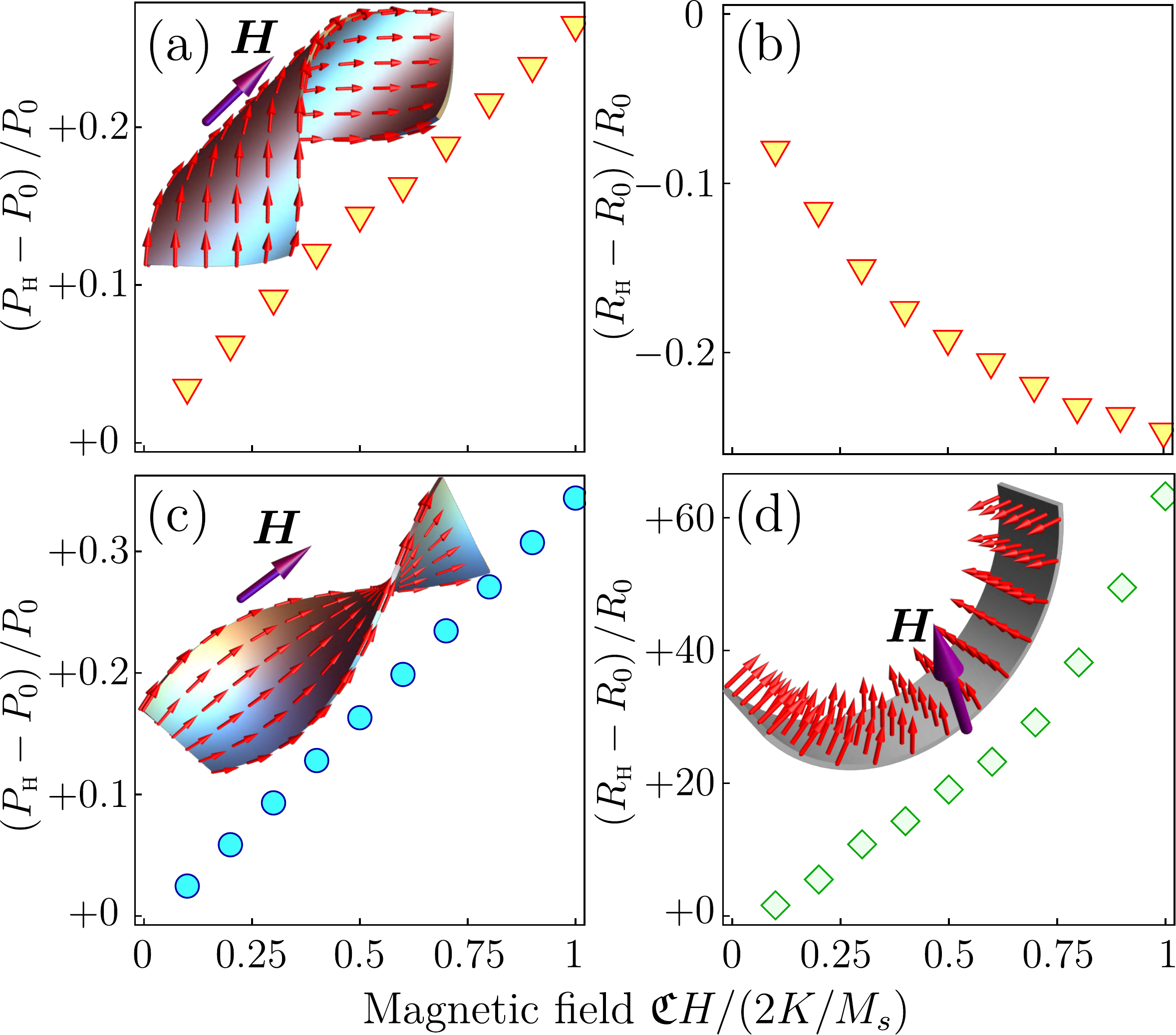}
	\caption{\label{fig:field}%
		(Color online) \textbf{Influence of the magnetic field on the geometrical parameters of the ribbon}: (a,b)  Relative field induced changes of radius and pitch of the DNA-like state as a function of applied magnetic field, respectively. (c) Relative field induced change of the pitch of the helicoid state as function of applied field. (d) Relative field induced change of the radius of the cylindrical state as function of applied field. All data are obtained by means of numerical simulations for ribbons with $w\approx2\ell$, $h=0.05\ell$, $A/\left(Yh^2\right)=1$, and $\nu=1/3$.  Insets demonstrate direction of the magnetic field $\vec{H}$ with respect to the deformed ribbon and magnetization orientation. Field induced dynamics is illustrated in the Supplemental movies~\ref{sec:video_bloch_field} and \ref{sec:video_neel_field}.}
\end{figure} 

For the case of rigid ribbon $\left(Y \to \infty\right)$ or vanishing DMI~$\left(D\to0\right)$, one gets values $P_0^\textsc{hel},\ P_0^\textsc{dna},\ R_0^\textsc{dna},\ R_0^\textsc{cyl}\to\infty$, which correspond to the straight ribbon.

\section{Influence of the external magnetic field}

Finally, we studied the influence of external magnetic field $\vec{H}$ on the equilibrium states considered above. The magnetic field was applied along the $\hat{\vec{z}}$ axis, i.e. $\vec{H}=H\hat{\vec{z}}$, for the DNA-like and helicoid states, while for the cylindrical state $\vec{H}=H\hat{\vec{x}}$. The interaction with magnetic field is represented by the Zeeman term with energy density $\mathcal{E}_\textsc{z}=-M_s\vec{m}\cdot\vec{H}$. The influence of the magnetic field was studied by means of the numerical simulations. 

Typical values of filed-induced changes of radii and pitches are presented in Fig.~\ref{fig:field}. One should note, that the relative field-induced deformations can reach up to 25--30~\% for the DNA-like and helicoid states, see Fig.~\ref{fig:field}(a-c), while for the case of cylindrical state it can reach up to~$\sim 10^3$ \%, see Fig.~\ref{fig:field}(d).

\section{Conclusions}

In conclusion, we predict the effect of spontaneous deformation of flexible ferromagnetic ribbon induced by DMI. The type  of deformation depends on the DMI symmetry and equilibrium magnetization distribution, see Tabl.~\ref{tab:states}. For DMI of Bloch type the deformation is possible only for the tangential magnetization of the ribbon and it is determined by the geometrical, mechanical, and magnetic parameters: DNA-like state  takes place for wide ribbons or relatively large $D$, while helicoid state is typical for narrow ribbons or relatively small $D$, see Fig.~\ref{fig:states}. In both cases the geometrical chirality of the ribbon is determined by the sign of $D$ and does not depend on the magnetization orientation along the ribbon. For the case of the N{\'e}el type DMI there is only one deformed state, namely the cylindrical state~(limit case of DNA-like state with $\psi=0$). It takes place only for the ribbons magnetized in normal direction. Finally, we show that geometrical parameters of the ribbon are significantly influenced by the external magnetic field, see Fig.~\ref{fig:field}. This feature can be used for control of the nanorobots mechanics.

\section{Conclusions}

We thank  U. Nitzsche for technical support and U. R{\"o}{\ss}ler for the helpful discussions. K.V.Y. acknowledges financial support from UKRATOP-project (funded by BMBF under reference 01DK18002). In part, this work was supported by the Program of Fundamental Research of the Department of Physics and Astronomy of the National Academy of Sciences of Ukraine (Project No.~0116U003192) and by Taras Shevcheko National University of Kyiv~(Project 19BF052-01).

\appendix
\begin{widetext}

\section{The model of flexible ferromagnetic ribbon}\label{sec:model}

The parametrization $\vec{\varsigma}(\xi_1,\xi_2)$ induces the natural tangential basis $\vec{g}_\alpha = \partial_\alpha\vec{\varsigma}$ with the corresponding metric tensor elements $g_{\alpha\beta}=\vec{g}_\alpha\cdot\vec{g}_\beta$. Here, $\alpha,\beta=1,2$ and $\partial_\alpha\equiv\partial_{\xi_\alpha}$. Assuming that vectors $\vec{g}_\alpha$ are orthogonal, one can introduce the orthonormal basis $\{\vec{e}_1, \vec{e}_2, \vec{n}\}$ with
\begin{equation}\label{eq:basis_suppl}
\vec{e}_\alpha=\frac{\vec{g}_\alpha}{\sqrt{g_{\alpha\alpha}}},\quad \vec{n}=\vec{e}_1\times\vec{e}_2.
\end{equation}
Using the Gau{\ss}-Godazzi formula and Weingarten's equation~\cite{Dubrovin84p1} one can obtain the following differential properties of the basis vectors
\begin{equation}\label{eq:basis_dif_suppl}
\nabla_\alpha\vec{e}_\beta=h_{\alpha\beta}\vec{n}-\Omega_{\alpha}\epsilon_{\beta\gamma}\vec{e}_\gamma,\quad \nabla_\alpha\vec{n}=-h_{\alpha\beta}\vec{e}_\beta.
\end{equation}
Here, $ \nabla_\alpha\equiv\left(g_{\alpha\alpha}\right)^{-1/2}\partial_\alpha$ (no summation over $\alpha$) are components of the curvilinear gradient and $\|h_{\alpha\beta}\|=\|b_{\alpha\beta}/\sqrt{g_{\alpha\alpha}g_{\beta\beta}}\|$ is modified second fundamental form. The second fundamental form determines the Gau{\ss} curvature $\mathcal{K}=\det\|h_{\alpha\beta}\|$ and mean curvature $\mathcal{H}=\text{tr}\|h_{\alpha\beta}\|$. Components of the spin connection vector $\vec{\Omega}$ are determined by the relation $\Omega_{\gamma}=\dfrac{1}{2}\epsilon_{\alpha\beta}\vec{e}_{\alpha}\cdot\nabla_\gamma\vec{e}_\beta$.

Using curvilinear reference frame~\eqref{eq:basis_suppl}, we introduce the following magnetization parametrization
\begin{equation}\label{eq:magnetization_suppl}
\vec{m}=\sin\theta\,\vec{\varepsilon}+\cos\theta\,\vec{n}, \quad \vec{\varepsilon}=\cos\phi\,\vec{e}_1+\sin\phi\,\vec{e}_2,
\end{equation}
where $\theta$ and $\phi$ are magnetic angles, and $\vec{\varepsilon}$ is a normalized projection of the vector $\vec{m}$ on the tangential plane.

We consider the flexible ferromagnetic ribbon with the following energy functional $E=E_\textsc{m}+E_\textsc{e}$, where
\begin{subequations}\label{eq:model_suppl}
	\begin{equation}\label{eq:model_m_suppl}
	E_\textsc{m}=h\int\left(A\mathcal{E}_\textsc{ex}+K\mathcal{E}_\textsc{a}+D\mathcal{E}_\textsc{d}\right)\sqrt{g}\textrm{d}\xi_1\mathrm{d}\xi_2
	\end{equation}
	is a magnetic energy term in total energy while
	\begin{equation}\label{eq:model_e_suppl}
	E_\textsc{e}=\frac{Yh}{8(1+\nu)}\int\left(\mathcal{E}_\textsc{s}+\frac{h^2}{3}\mathcal{E}_\textsc{b}\right)\sqrt{\overline{g}}\textrm{d}\xi_1\mathrm{d}\xi_2,
	\end{equation}
	is an elastic energy term. Elastic energy taken for the thin ribbons where only terms of first and third order of magnitude with respect to thickness $h$ are taken into account~\cite{Efrati09}. Here, $g=\det\|g_{\alpha\beta}\|$, and $\overline{g}=\det\|\overline{g}_{\alpha\beta}\|$	with $\overline{g}_{\alpha\beta}$ being metric tensor for ribbon free of elastic tensions (we consider $\overline{g}_{\alpha\beta}=\delta_{\alpha\beta}$). Parameters $Y$ and $\nu\in\left[0,0.5\right]$ in \eqref{eq:model_e_suppl} are Young's modulus and Poisson ratio, respectively.
\end{subequations}

The first term in \eqref{eq:model_m_suppl} is the exchange density $\mathcal{E}_\textsc{ex}=\sum_{i=x,y,z}\left(\partial_i\vec{m}\right)^2$ with $A$ being exchange constant. In the curvilinear reference frame exchange energy can be written as~\cite{Gaididei14,Sheka15,Kravchuk16a} 
\begin{subequations}\label{eq:exchnage_suppl}
	\begin{equation}\label{eq:exchnage_m_suppl}
	\begin{split}
	\mathcal{E}_\textsc{ex}=&\nabla_\alpha m_\beta\nabla_\alpha m_\beta+\nabla_\alpha m_n\nabla_\alpha m_n\\
	+ &2h_{\alpha\beta}\left(m_\beta\nabla_\alpha m_n - m_n\nabla_\alpha m_\beta\right)+2\epsilon_{\alpha\beta}\Omega_\gamma m_\beta\nabla_\gamma m_\alpha\\
	+ &\left(h_{\alpha\gamma}h_{\gamma\beta}+\Omega^2\delta_{\alpha\beta}\right)m_\alpha m_\beta+\left(\mathcal{H}^2-2\mathcal{K}\right)m_n^2+2\epsilon_{\alpha\gamma}h_{\gamma\beta}\Omega_{\beta}m_\alpha m_n.
	\end{split}
	\end{equation}
	Using the angular parametrization~\eqref{eq:magnetization_suppl} one can obtain~\cite{Gaididei14,Sheka15,Kravchuk16a} 
	\begin{equation}\label{eq:exchnage_angle_suppl}
	\mathcal{E}_\textsc{ex}=\left[\vec{\nabla}\theta-\vec{\Gamma}\right]^2+\left[\sin\theta\left(\vec{\nabla}\phi-\vec{\Omega}\right)-\cos\theta\partial_\phi\vec{\Gamma}\right],
	\end{equation}
	where $\vec{\Gamma}=\|h_{\alpha\beta}\|\cdot\vec{\varepsilon}$.
\end{subequations}

The second term in the~\eqref{eq:model_m_suppl} correspond to the Dzyaloshinskii--Moriya interaction~(DMI) $\mathcal{E}_\textsc{d}$, with $D$ being DMI constant. For the case of a N{\'e}el type DMI $\mathcal{E}_\textsc{d}^\textsc{n}=m_n\vec{\nabla}\cdot\vec{m}-\vec{m}\cdot\vec{\nabla}m_n$ in the curvilinear reference frame this interaction can be written as~\cite{Kravchuk16a}
\begin{subequations}\label{eq:dmi_suppl}
	\begin{equation}\label{eq:dmi_surface_suppl}
	\begin{split}
	\mathcal{E}_\textsc{d}^\textsc{n}=m_n\nabla_\alpha m_\alpha-m_\alpha\nabla_\alpha m_n-\epsilon_{\alpha\beta}\Omega_{\beta}m_\alpha m_n - \mathcal{H} m_n^2.
	\end{split}
	\end{equation}
	Using the angular parametrization~\eqref{eq:magnetization_suppl} one can obtain~\cite{Kravchuk16a,Kravchuk18a} 
	\begin{equation}\label{eq:dmi_surface_angle_suppl}
	\mathcal{E}_\textsc{d}^\textsc{n}=2\left(\vec{\nabla}\theta\cdot\vec{\varepsilon}\right)\sin^2\theta-\mathcal{H}\cos^2\theta,
	\end{equation}
	While, for the Bloch type DMI symmetry $\mathcal{E}_\textsc{d}^\textsc{b}=\vec{m}\cdot\left[\vec{\nabla}\times\vec{m}\right]$ this interaction in curvilinear reference frame reads as 
	\begin{equation}\label{eq:dmi_bulk_suppl}
	\begin{split}
	\mathcal{E}_\textsc{d}^\textsc{b}=\epsilon_{\alpha\beta}\left(m_n\nabla_\alpha m_\beta-m_\beta\nabla_\alpha m_n\right)+\epsilon_{\alpha\beta}h_{\beta\gamma}m_\alpha m_\gamma - \Omega_\alpha m_\alpha m_n.
	\end{split}
	\end{equation}
	Substituting the angular parametrization~\eqref{eq:magnetization_suppl} into~\eqref{eq:dmi_bulk_suppl} results in the expression (up to the boundary terms)
	\begin{equation}\label{eq:dmi_bulk_angle_suppl}
	\mathcal{E}_\textsc{d}^\textsc{b} = \sin^2\theta\,\left[\left(2\vec{\nabla}\theta-\vec{\Gamma}\right)\times\vec{\varepsilon}\right]\cdot\vec{n}.
	\end{equation}
\end{subequations}

Last term in~\eqref{eq:model_m_suppl} corresponds to the uniaxial anisotropy $\mathcal{E}_\textsc{a}=1-\left(\vec{m}\cdot\vec{e}_\textsc{a}\right)^2$, with $K>0$ being easy-axial anisotropy constant. In a curvilinear reference frame anisotropy contribution has particularly simple form
\begin{equation}\label{eq:anisotropy_suppl}
\begin{split}
\mathcal{E}_\textsc{a}^\textsc{tang}=&1-\sin^2\theta\cos^2\phi,\ \text{where } \vec{e}_\textsc{a}=\vec{e}_1,\\
\mathcal{E}_\textsc{a}^\textsc{norm}=&\sin^2\theta,\ \text{where } \vec{e}_\textsc{a}=\vec{n}.
\end{split}
\end{equation}

The first term in elastic energy~\eqref{eq:model_e_suppl} corresponds to the stretching energy~\cite{Efrati09}
\begin{equation}\label{eq:stretching_suppl}
\mathcal{E}_\textsc{s}=\left(\frac{\nu}{1-\nu}\overline{g}^{\alpha\beta}\overline{g}^{\gamma\delta}+\overline{g}^{\alpha\gamma}\overline{g}^{\beta\delta}\right)\left(g_{\alpha\beta}-\overline{g}_{\alpha\beta}\right)\left(g_{\gamma\delta}-\overline{g}_{\gamma\delta}\right).
\end{equation}
The last term in~\eqref{eq:model_e_suppl} determines the bending energy~\cite{Efrati09}
\begin{equation}\label{eq:bending_suppl}
\mathcal{E}_\textsc{b}=\left(\frac{\nu}{1-\nu}\overline{g}^{\alpha\beta}\overline{g}^{\gamma\delta}+\overline{g}^{\alpha\gamma}\overline{g}^{\beta\delta}\right)b_{\alpha\beta}b_{\gamma\delta}.
\end{equation}

In our analytical calculations we used two different geometry parametrizations: DNA-like and helicoid geometries. 

\subsection{Geometrical properties and elastic energy of the DNA-like deformation}

DNA-like geometry can be parametrized in the following way~[see Fig.~\ref{fig:states}(a)]
\begin{equation}\label{eq:dna_parametrization_suppl}
\vec{\varsigma}^\textsc{dna}\left(\xi_1,\xi_2\right)=R\cos\left(\frac{\rho}{R}\right)\hat{\vec{x}}+R\sin\left(\frac{\rho}{R}\right)\hat{\vec{y}}+\left(\xi_1\sin\psi+\xi_2\cos\psi\right)\hat{\vec{z}}, \quad \rho=\xi_1\cos\psi-\xi_2\sin\psi,
\end{equation}
where $R$ is a radius of the central line, and $\psi$ is an angle between vector $\vec{e}_1$ and $\vec{xy}$-plane in the tangential plane [see Fig.~\ref{fig:states}(a)]. The pitch of the DNA-like state defined as $P=2\pi R\tan\psi$.

Parametrization \eqref{eq:dna_parametrization_suppl} results in the following first, second, and modified second fundamental forms
\begin{equation}\label{eq:dna_1_2_forms_suppl}
g_{\alpha\beta}=\delta_{\alpha\beta},\quad \|b_{\alpha\beta}\|=\|h_{\alpha\beta}\|=\frac{1}{R}\left\|\begin{matrix}
-\cos^2\psi&\cos\psi\sin\psi\\
\cos\psi\sin\psi&-\sin^2\psi
\end{matrix}\right\|,
\end{equation}
respectively. DNA-like geometry has zero Gau{\ss} curvature $\mathcal{K}=0$, nonzero mean curvature $\mathcal{H}=-1/R$ (here minus is related to the direction of the normal vector), and zero components of spin connection vector $\vec{\Omega}=\vec{0}$.

Using definitions of first and second fundamental forms for DNA-like deformation~\eqref{eq:dna_1_2_forms_suppl} one can obtain expressions for elastic energy densities \eqref{eq:stretching_suppl} and \eqref{eq:bending_suppl} in form
\begin{equation}\label{eq:dna_elastic_suppl}
\mathcal{E}_\textsc{s} = 0, \qquad \mathcal{E}_\textsc{b}=\frac{1}{\left(1-\nu\right)R^2}.
\end{equation}
From \eqref{eq:dna_elastic_suppl} one can see that DNA-like deformations are free from stretching energy.

\subsection{Geometrical properties and elastic energy of the helicoid deformation}

Helicoid geometry can be parametrized in the following way~[see Fig.~\ref{fig:states}(b)]
\begin{equation}\label{eq:helicoid_parametrization_suppl}
\vec{\varsigma}^\textsc{hel}\left(\xi_1,\xi_2\right)=\xi_2\left[\cos\left(k\xi_1\right)\hat{\vec{x}}+\sin\left(k\xi_1\right)\hat{\vec{y}}\right]+\xi_1\hat{\vec{z}},
\end{equation}
where $k$ is a twist parameter, which results in the pitch $P=2\pi/k$. Parametrization~\eqref{eq:helicoid_parametrization_suppl} results in the following first, second, and modified second fundamental forms
\begin{equation}\label{eq:helicoid_1_2_forms_suppl}
\|g_{\alpha\beta}\|=\text{diag}\left(1+k^2\xi_2^2,1\right), \quad \|b_{\alpha\beta}\|=\frac{k}{\sqrt{1+\left(k\xi_2\right)^2}}\text{adiag}\left(1,1\right), \quad \|h_{\alpha\beta}\|=\frac{k}{1+\left(k\xi_2\right)^2}\text{adiag}\left(1,1\right),
\end{equation}
respectively. Helicoid geometry, in contrast to DNA-like~\eqref{eq:dna_parametrization_suppl}, has nonzero Gau{\ss} curvature $\mathcal{K}=-k^2/\left(1+k^2\xi_2^2\right)$, zero mean curvature $\mathcal{H}=0$, and nonzero component of spin connection vector $\vec{\Omega}=\dfrac{k^2\xi_2}{\left(1+k^2\xi_2^2\right)^2}\vec{e}_1$.

Using definitions of first and second fundamental forms for helicoid deformation~\eqref{eq:helicoid_1_2_forms_suppl} one can obtain expressions for elastic energy densities \eqref{eq:stretching_suppl} and \eqref{eq:bending_suppl} in form
\begin{equation}\label{eq:helicoid_elastic_suppl}
\mathcal{E}_\textsc{s} = \frac{1}{1-\nu}k^4\xi_2^4, \qquad \mathcal{E}_\textsc{b}=\frac{2k^2}{1+k^2\xi^2_2}.
\end{equation}

\section{DMI of Bloch type}\label{sec:bulk_dmi}

In this section we consider DMI in form  $\mathcal{E}_\textsc{d}=\mathcal{E}_\textsc{d}^\textsc{b}$ which is defined in Eqs.~\eqref{eq:dmi_bulk_suppl} and~\eqref{eq:dmi_bulk_angle_suppl}. Here we also will consider two different easy-axial anisotropy directions.

\subsection{DNA-like geometry}\label{sec:dna_state_bulk}
\subsubsection{Case of easy-tangential anisotropy}\label{sec:dna_bulk_tangential}
We start with a DNA-like geometry defined in \eqref{eq:dna_parametrization_suppl} with easy-tangential anisotropy~($\vec{e}_\textsc{a}=\vec{e}_1$). We are interested in the equilibrium states. Therefore, we consider the simplest case when magnetization is uniform in the curvilinear reference frame, i.e. $\theta=\text{const}$ and $\phi=\text{const}$. The total energy~\eqref{eq:model_suppl} reads as
\begin{equation}\label{eq:dna_energy_suppl}
\frac{E^\textsc{dna}}{hwL}=\frac{A}{R^2}\left[1-\sin^2\theta\sin^2\left(\phi+\psi\right)\right]+\frac{D}{R}\sin^2\theta\sin\left(\phi+\psi\right)\cos\left(\phi+\psi\right)+K\left(1-\sin^2\theta\cos^2\phi\right)+\frac{Yh^2}{24R^2\left(1-\nu^2\right)}.
\end{equation}  

By minimization of~\eqref{eq:dna_energy_suppl} we obtain that magnetic angles are defined as $\theta_0^\textsc{dna}=\pi/2$ and $\cos\phi^\textsc{dna}_0=\mathfrak{C}$. Here $\mathfrak{C}=\pm1$ determines whether magnetization is parallel~($\mathfrak{C}=+1$) or antiparallel ($\mathfrak{C}=-1$) to the tangential axis~(i.e. $\vec{m}_0^\textsc{dna}=\mathfrak{C}\, \vec{e}_1$). The equilibrium values of the radius $R$ and angle $\psi$ is defined by the following equations
\begin{equation}\label{eq:dna_r_psi_suppl}
\frac{D}{R}\cos2\psi-\frac{A}{R^2}\sin2\psi=0,\quad\frac{Yh^2}{12R^3\left(1-\nu^2\right)}+\frac{A}{R^3}\left(1+\cos2\psi\right)+\frac{D}{2R^2}\sin2\psi=0.
\end{equation}
Equations~\eqref{eq:dna_r_psi_suppl} results in the following equilibrium geometrical parameters
\begin{equation}\label{eq:dna_geo_equilibrium_suppl}
\begin{split}
R_0^\textsc{dna}=&\frac{A}{|D|}\frac{2\sqrt[4]{1+\zeta}}{\sqrt{1+\zeta}-1},\quad \cos\psi_0^\textsc{dna}=-\frac{\text{sgn}\left(D\right)}{\sqrt{1+\sqrt{1+\zeta}}},\\ P_0^\textsc{dna}=&\frac{A}{D} \frac{4\pi \sqrt{1+\zeta}}{\sqrt{1+\zeta}-1},\quad\zeta=24\left(1-\nu^2\right)\frac{A}{Yh^2}.
\end{split}
\end{equation}
The energy of the DNA-like state reads
\begin{equation}\label{eq:dna_energy_equlibrium_suppl}
\frac{E_0^\textsc{dna}}{hwL}=-\frac{D^2}{4A}\frac{\sqrt{1+\zeta}-1}{\sqrt{1+\zeta}+1}.
\end{equation}

For the case of large values of Young's modulus $\left(\frac{A}{Yh^2}\ll1\right)$, equilibrium parameters~\eqref{eq:dna_geo_equilibrium_suppl} can be written as
\begin{equation}\label{eq:dna_geo_approx_suppl}
\begin{split}
R_0^\textsc{dna}\approx&\frac{1}{6\left(1-\nu^2\right)}\frac{Yh^2}{|D|}\left[1+12\left(1-\nu^2\right)\frac{A}{Yh^2}\right]-9\left(1-\nu^2\right)\frac{A}{|D|}\frac{A}{Yh^2},\\
P_0^\textsc{dna}\approx&\frac{\pi}{3\left(1-\nu^2\right)}\frac{Yh^2}{D}\left[1+18\left(1-\nu^2\right)\frac{A}{Yh^2}\right]-12\pi\left(1-\nu^2\right)\frac{A}{D}\frac{A}{Yh^2},\\
\cos\psi_0^\textsc{dna}\approx&-\frac{\text{sgn}(D)}{\sqrt{2}}\left[1-3\left(1-\nu^2\right)\frac{A}{Yh^2}\right],
\end{split}
\end{equation}
with energy
\begin{equation}\label{eq:dna_energy_approx_suppl}
\frac{E_0^\textsc{dna}}{hwL}\approx -3\left(1-\nu^2\right)\frac{D^2}{Yh^2}\left[1-6\left(1-\nu^2\right)\frac{A}{Yh^2}\right].
\end{equation}

\subsubsection{Case of easy-normal anisotropy}\label{sec:dna_bulk_normal}

For the case of easy-normal anisotropy~($\vec{e}_\textsc{a}=\vec{n}$) it is convenient to introduce new parametrization for the unit magnetization vector
\begin{equation}\label{eq:magnetization_1_suppl}
\vec{m}=\cos\Theta\,\vec{e}_1+\sin\Theta\cos\Phi\,\vec{e}_2+\sin\Theta\sin\Phi\,\vec{n}.
\end{equation}
Therefore, using parametrization~\eqref{eq:magnetization_1_suppl} one can write the total energy~\eqref{eq:model_suppl} in form (for the case of uniform magnetization distribution in curvilinear reference frame)
\begin{equation}\label{eq:dna_bulk_norm_suppl}
\begin{split}
\frac{E^\textsc{dna}}{hwL}=\frac{A}{R^2}&\left[\sin^2\Theta+\cos^2\psi\left(\cos^2\Theta-\sin^2\Theta\sin^2\Phi\right)-\frac{1}{2}\sin2\Theta\sin2\psi\sin\Phi\right]\\
+\frac{D}{2R}&\left[\sin 2\Theta\cos2\psi\cos\Phi+\sin 2\psi\left(\cos^2\Theta-\sin^2\Theta\cos^2\Phi\right)\right]+K\left(1-\sin^2\Theta\sin^2\Phi\right)+\frac{Yh^2}{24R^2\left(1-\nu^2\right)}.
\end{split}
\end{equation}
By minimization of~\eqref{eq:dna_bulk_norm_suppl} we obtain that magnetic angles are defined as $\Theta_0^\textsc{dna}=\pm \pi/2$ and $\Phi_0^\textsc{dna}=\pm \pi/2$, i.e. $\vec{m}_0^\textsc{dna}=\pm\vec{n}$. For such magnetization distribution the DMI energy is equal to zero~$E_\textsc{d} = 0$. The letter effect results in not deformed ribbon with equilibrium values for the geometrical parameters results in $\sin\psi_0^\textsc{dna}=0$ and $R\to\infty$, which correspond to the straight and not deformed ribbon. Total energy of this state~\eqref{eq:dna_bulk_norm_suppl}  is zero $E^\textsc{dna}\left(\vec{e}_\textsc{a}=\vec{n}\right)=0$.

\subsection{Helicoid geometry}\label{sec:hel_state}
\subsubsection{Case of easy-tangential anisotropy}\label{sec:hel_bulk_tangential}
Here, we consider a helicoid geometry~\eqref{eq:helicoid_parametrization_suppl} with easy-tangential anisotropy. Similarly to the DNA-like geometry discussed in~\ref{sec:dna_state_bulk}, here, we are interested in the equilibrium states with uniform magnetization in the curvilinear reference frame. Therefore, the total energy~\eqref{eq:model_suppl} can be written as
\begin{equation}\label{eq:hel_energy_suppl}
\begin{split}
\frac{E^\textsc{hel}}{hwL}=2A\frac{k}{w}\left[\frac{2kw\cos^2\theta}{\sqrt{4+\left(kw\right)^2}}+\text{arcsinh}\left(\frac{kw}{2}\right)\sin^2\theta\right]+2\frac{D}{w}\text{arcsinh}\left(\frac{kw}{2}\right)\sin^2\theta\cos2\phi+K\left(1-\sin^2\theta\cos^2\phi\right)\\
+\frac{Y}{640\left(1-\nu^2\right)}\left(kw\right)^4+\frac{Yh^2}{6\left(1+\nu\right)}\frac{k}{w}\arctan\left(\frac{kw}{2}\right).
\end{split}
\end{equation}
We consider narrow ribbons and we assume that DMI induced deformations are small $kw\ll1$. Therefore, we will write energy of the helicoid state as
\begin{equation}\label{eq:hel_energ_approx_suppl}
\begin{split}
\frac{E^\textsc{hel}}{hwL}\approx A k^2\left[1+\cos^2\theta-\frac{k^2w^2}{24}\left(1+5\cos^2\theta\right)\right]+D& k\left(1-\frac{k^2w^2}{24}\right)\sin^2\theta\cos2\phi\\
-K\sin^2\theta\cos^2\phi+\frac{Yh^2k^2}{12\left(1+\nu\right)}&+\frac{Y\left(kw\right)^4}{640\left(1-\nu^2\right)},
\end{split}
\end{equation}
where we save terms up to fourth order of magnitude with respect to $kw$.


Energy~\eqref{eq:hel_energ_approx_suppl} has minimum for magnetic angles $\theta_0^\textsc{hel}=\pi/2$ and $\cos\phi^\textsc{hel}_0=\mathfrak{C}$. The equilibrium value of the twist parameter $k$ is defined by the following equation
\begin{equation}\label{eq:hel_param_def_suppl}
D\left(1-\frac{k^2w^2}{8}\right)+2Ak\left(1-\frac{k^2w^2}{12}\right)+\frac{Yh^2k}{6\left(1+\nu\right)}\left(1-\frac{k^2w^2}{6}\right)+\frac{Yw^4k^3}{160\left(1-\nu^2\right)}=0.
\end{equation}
The solution of equation~\eqref{eq:hel_param_def_suppl} coincides with the twist parameter which obtained from numerical minimization of energy~\eqref{eq:hel_energy_suppl} with an accuracy of about $4\times 10^{-3}$~(for Young's modulus in the range $A/\left(Yh^2\right)\in\left[0,1\right]$). 

For the case of relatively large Young's modulus ($A/\left(Yh^2\right)\ll1$) the twist parameter and the pitch of helicoid state can be defined as
\begin{equation}
k_0^\textsc{hel}\approx - 6\frac{D}{Yh^2}\frac{1+\nu}{1+12\left(1+\nu\right)A/\left(Yh^2\right)},\quad P_0^\textsc{hel} = 2\pi/k_0^\textsc{hel}.
\end{equation}

The energy of a helicoid state approximately can be determined as
\begin{equation}
\frac{E_0^\textsc{hel}}{hwL}\approx-3\left(1+\nu\right)\frac{D^2 }{Yh^2}\biggl[1-\frac{27}{40}\frac{D^2}{Y^2h^2}\frac{w^4}{h^4}\frac{\left(1+\nu\right)^2}{\left(1-\nu\right)}\biggr].
\end{equation}

By comparing energies of DNA-like state~\eqref{eq:dna_energy_equlibrium_suppl} and helical state~\eqref{eq:hel_energ_approx_suppl} [see Fig.~\ref{fig:energy}(a)] for the easy-tangential anisotropy and twist parameter from~\eqref{eq:hel_param_def_suppl}, we find the energetically preferable states for different DMI constants and Young's modulus values. The resulting phase diagrams are presented in Fig.~\ref{fig:states}(c)-(d).

\begin{figure}
	\includegraphics[width=0.75\textwidth]{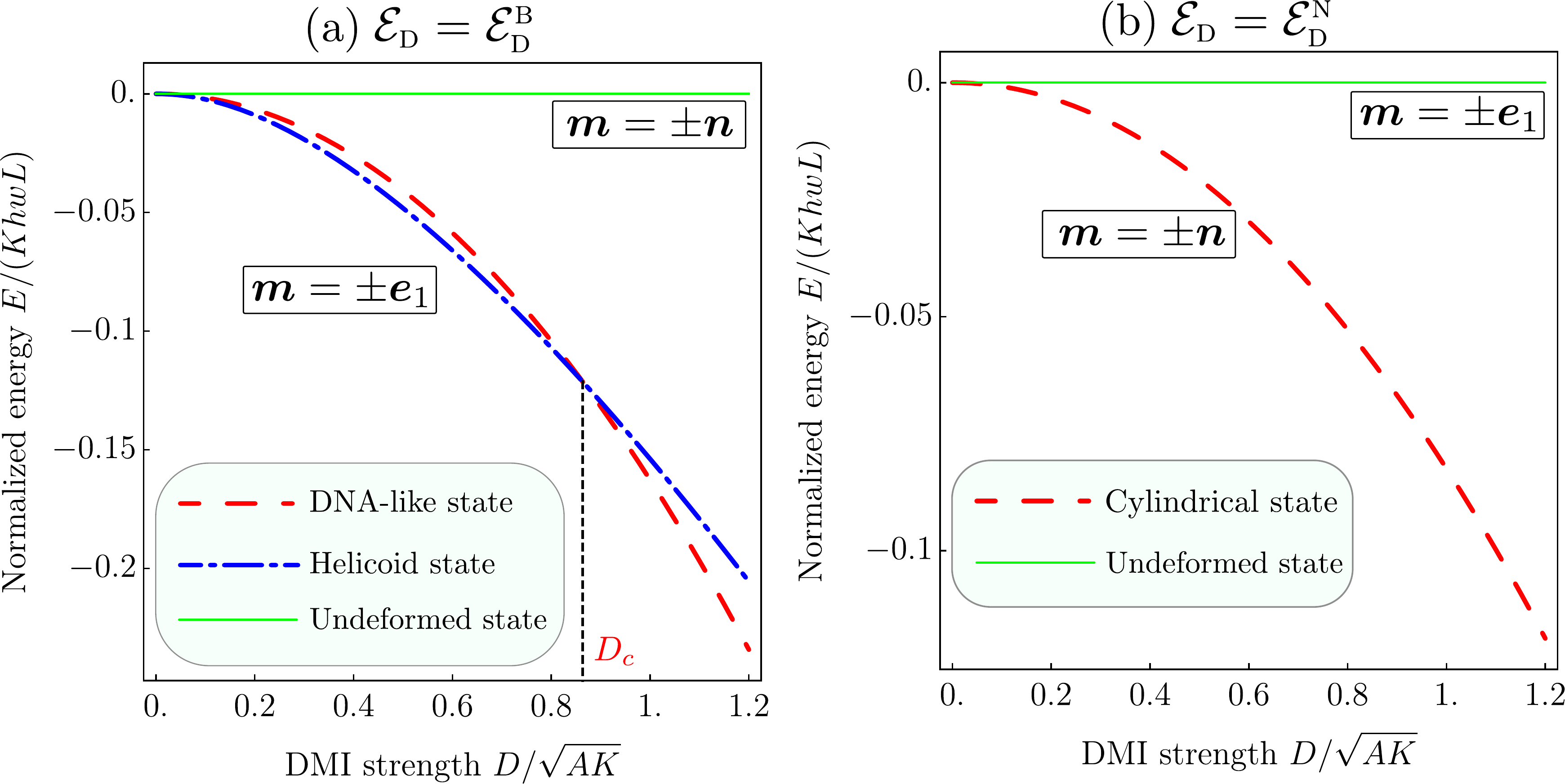}
	\caption{\label{fig:energy}%
		(Color online) \textbf{Energy of the flexible ribbon}: (a) Energies of flexible ribbon in DNA-like, helicoid, and undeformed states for Bloch type DMI. (b) Energies of flexible ribbon in cylindrical and undeformed states for N{\'e}el type DMI. In all cases $\nu=1/3$, $h=0.1\ell$, $w=2.5\ell$, $A/\left(Yh^2\right)=1$.}
\end{figure} 

\subsubsection{Case of easy-normal anisotropy}\label{sec:hel_bulk_normal}

Here we will use magnetization parametrization~\eqref{eq:magnetization_1_suppl} defined in Sec.~\ref{sec:dna_bulk_normal}. The energy for the helicoid geometry with easy-normal anisotropy can be written as
\begin{equation}\label{eq:hel_bulk_norm_suppl}
\begin{split}
\frac{E^\textsc{hel}}{hwL}=2A\frac{k}{w}\biggl[\frac{2kw}{\sqrt{4+k^2w^2}}\sin^2\Theta\sin^2\Phi+\text{arcsinh}\frac{kw}{2}\left(\cos^2\Theta+\sin^2\Theta\cos^2\Phi\right)\biggr]+K\left(1-\sin^2\Theta\sin^2\Phi\right)\\
+2\frac{D}{w}\text{arcsinh}\frac{kw}{2}\left(\cos^2\Theta-\sin^2\Theta\cos^2\Phi\right)+\frac{Y}{640\left(1-\nu^2\right)}\left(kw\right)^4+\frac{Yh^2}{6\left(1+\nu\right)}\frac{k}{w}\arctan\left(\frac{kw}{2}\right).
\end{split}
\end{equation}

Energy~\eqref{eq:hel_bulk_norm_suppl} has minimum for magnetic angles $\Theta_0^\textsc{hel}=\pm\pi/2$ and $\Phi^\textsc{hel}_0=\pm \pi/2$ which results in zero DMI energy~$E_\textsc{d} = 0$. The latter effect results in not deformed ribbon with twist parameter $k=0$.  Total energy of this state~\eqref{eq:hel_bulk_norm_suppl}  is zero $E^\textsc{hel}\left(\vec{e}_\textsc{a}=\vec{n}\right)=0$.

\section{DMI of N{\'e}el type}\label{sec:interfacial_dmi}

In this section we consider interfacial DMI $\mathcal{E}_\textsc{d}=\mathcal{E}_\textsc{d}^\textsc{n}$ defined in Eqs.~\eqref{eq:dmi_surface_suppl} and \eqref{eq:dmi_surface_angle_suppl}. Here we will perform the same calculations as described in Sec.~\ref{sec:bulk_dmi}.

\subsection{DNA-like geometry}
\subsubsection{Case of easy-tangential anisotropy}\label{sec:dna_interfacial_tangential}
Here we consider a DNA-like geometry defined in \eqref{eq:dna_parametrization_suppl} with easy-tangential anisotropy~($\vec{e}_\textsc{a}=\vec{e}_1$). For uniform  magnetization distribution in a curvilinear reference frame the energy of the ribbons reads
\begin{equation}\label{eq:dna_interfacial_energy_suppl}
\frac{E^\textsc{dna}}{hwL}=\frac{A}{R^2}\left[1-\sin^2\theta\sin^2\left(\phi+\psi\right)\right]-\frac{D}{R}\cos^2\theta+K\left(1-\sin^2\theta\cos^2\phi\right)+\frac{Yh^2}{24R^2\left(1-\nu^2\right)}.
\end{equation}
Energy \eqref{eq:dna_interfacial_energy_suppl} has minimum for magnetic angles $\theta_0^\textsc{dna}=\pm\pi/2$ and $\cos\phi^\textsc{dna}_0=\mathfrak{C}$~(magnetization is align along the ribbon, i.e. $\vec{m}^\textsc{dna}_0=\mathfrak{C}\vec{e}_1$), which results in zero DMI energy~$E_\textsc{d} = 0$. The latter effect results in not deformed ribbon with geometrical parameters $\psi_0^\textsc{dna}=\pi/2$ and $R\to\infty$.   Total energy of this state~\eqref{eq:dna_interfacial_energy_suppl}  is zero $E^\textsc{dna}\left(\vec{e}_\textsc{a}=\vec{e}_1\right)=0$.

\subsubsection{Case of easy-normal anisotropy}\label{sec:dna_interfacial_normal}
Here we will use magnetization parametrization~\eqref{eq:magnetization_1_suppl}. The energy~\eqref{eq:model_suppl} for the DNA-like geometry with easy-normal anisotropy can be written as
\begin{equation}\label{eq:dna_interfacial_energy_normal_suppl}
\begin{split}
\frac{E^\textsc{dna}}{hwL}=\frac{A}{R^2}&\left[\sin^2\Theta+\cos^2\psi\left(\cos^2\Theta-\sin^2\Theta\sin^2\Phi\right)-\frac{1}{2}\sin2\Theta\sin2\psi\sin\Phi\right]\\
-\frac{D}{R}&\sin^2\Theta\sin^2\Phi+K\left(1-\sin^2\Theta\sin^2\Phi\right)+\frac{Yh^2}{24R^2\left(1-\nu^2\right)}.
\end{split}
\end{equation}
Energy~\eqref{eq:dna_interfacial_energy_normal_suppl} has minimum for magnetic angles $\Theta_{0}^\textsc{dna}=\pm \pi/2$ and $\Phi_{0}^\textsc{dna}=\pm\pi/2$ with $\sin\psi_{0}^\textsc{dna}=0$ and equilibrium radius
\begin{equation}\label{eq:radius_cylinder_suppl}
R_0^\textsc{cyl}=2\frac{A}{|D|}\left[1+\frac{Yh^2}{24A\left(1+\nu\right)}\right].
\end{equation}
This state has pitch equal to zero $P=0$, therfore we refer this state as cylindrical.

The energy of this state is 
\begin{equation}\label{eq:energy_cylinder_suppl}
\frac{E_0^\textsc{cyl}}{hwL}=-\frac{D^2}{Y h^2}\frac{6\left(1+\nu\right)}{1+24A\left(1+\nu\right)/\left(Yh^2\right)}.
\end{equation}
Fig.~\ref{fig:energy}(b) shows the comparison of energies of flexible ribbon with interfacial DMI and easy-normal anisotropy in DNA-like and helicoid states.
\begin{table}
	\begin{tabular}{p{0.05\textwidth}||p{0.125\textwidth}|p{0.375\textwidth}|p{0.45\textwidth}|}
		\centering DMI type & \centering Magnetization direction & \centering Equilibrium state & {\centering Equilibrium parameters} \\
		\hline
		\hline
		\multirow{3}{*}{\rotatebox{90}{\centering Bulk DMI: $\mathcal{E}_\textsc{d}^\textsc{b}=\vec{m}\cdot\left[\vec{\nabla}\times\vec{m}\right]$}} & \centering \multirow{2}{*}{\rotatebox{0}{ $\vec{m}=\pm\vec{e}_1$}}  & {\includegraphics[width=0.375\textwidth]{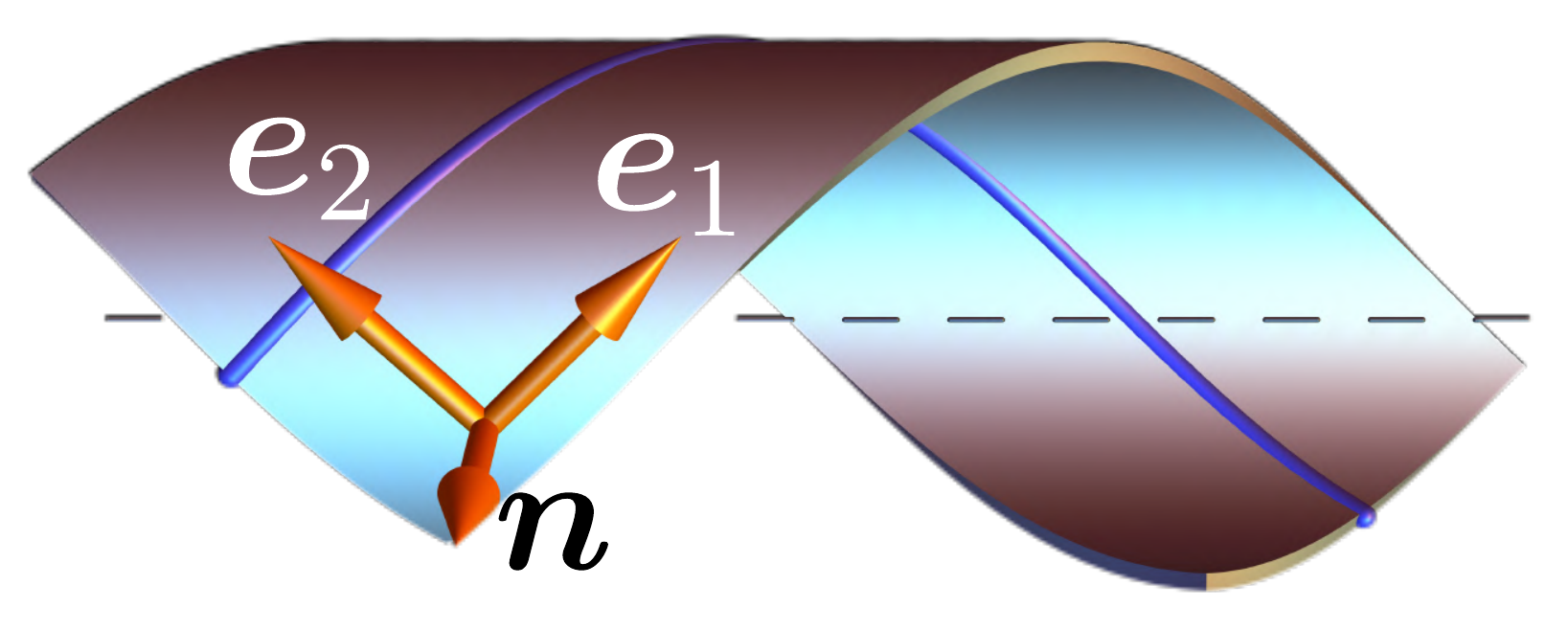}} &	\vspace{-2.5cm}\begin{itemize}
			\item $R_0^\textsc{dna} = \dfrac{A}{|D|}\dfrac{2\sqrt[4]{1+\zeta}}{\sqrt{1+\zeta} - 1}$;
			\item $P_0^\textsc{dna} = \dfrac{A}{D} \dfrac{4\pi \sqrt{1+\zeta}}{\sqrt{1+\zeta}-1}$;
			\item $\dfrac{E_0^\textsc{dna}}{hwL}=-\dfrac{D^2}{4A}\dfrac{\sqrt{1+\zeta}-1}{\sqrt{1+\zeta}+1}.$
		\end{itemize} \\
		\cline{3-4}
		&  & \includegraphics[width=0.375\textwidth]{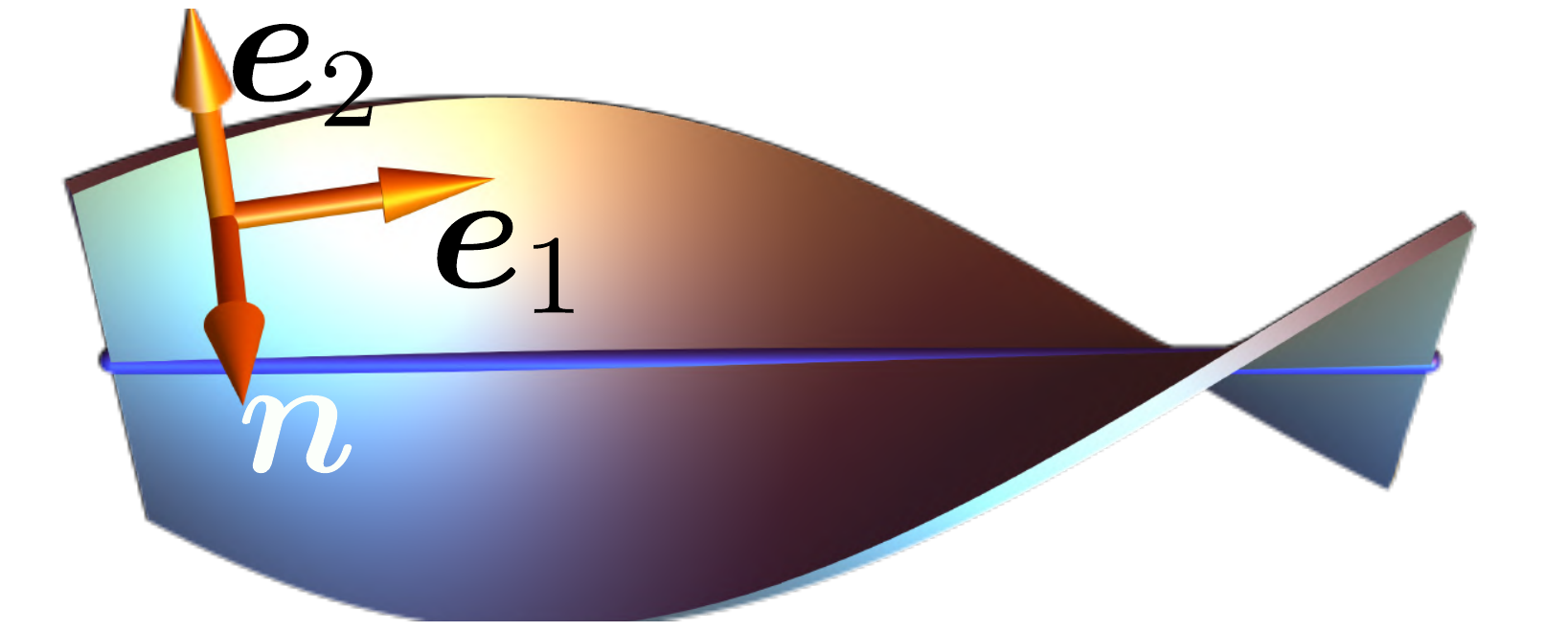} & \vspace{-2.5cm}\begin{itemize}
			\item $P_0^\textsc{hel}\approx\dfrac{\pi}{3\left(1+\nu\right)}\dfrac{Yh^2}{D}\left[1+12\left(1+\nu\right)\dfrac{A}{Yh^2}\right]$;
			\item $\dfrac{E_0^\textsc{hel}}{hwL}\approx-3\left(1+\nu\right)\dfrac{D^2 }{Yh^2}\biggl[1-\dfrac{27}{40}\dfrac{D^2}{Y^2h^2}\dfrac{w^4}{h^4}\dfrac{\left(1+\nu\right)^2}{\left(1-\nu\right)}\biggr]$.
		\end{itemize} \\
		\cline{2-4} 
		& \centering $\vec{m}=\pm\vec{n}$ & \includegraphics[width=0.375\textwidth]{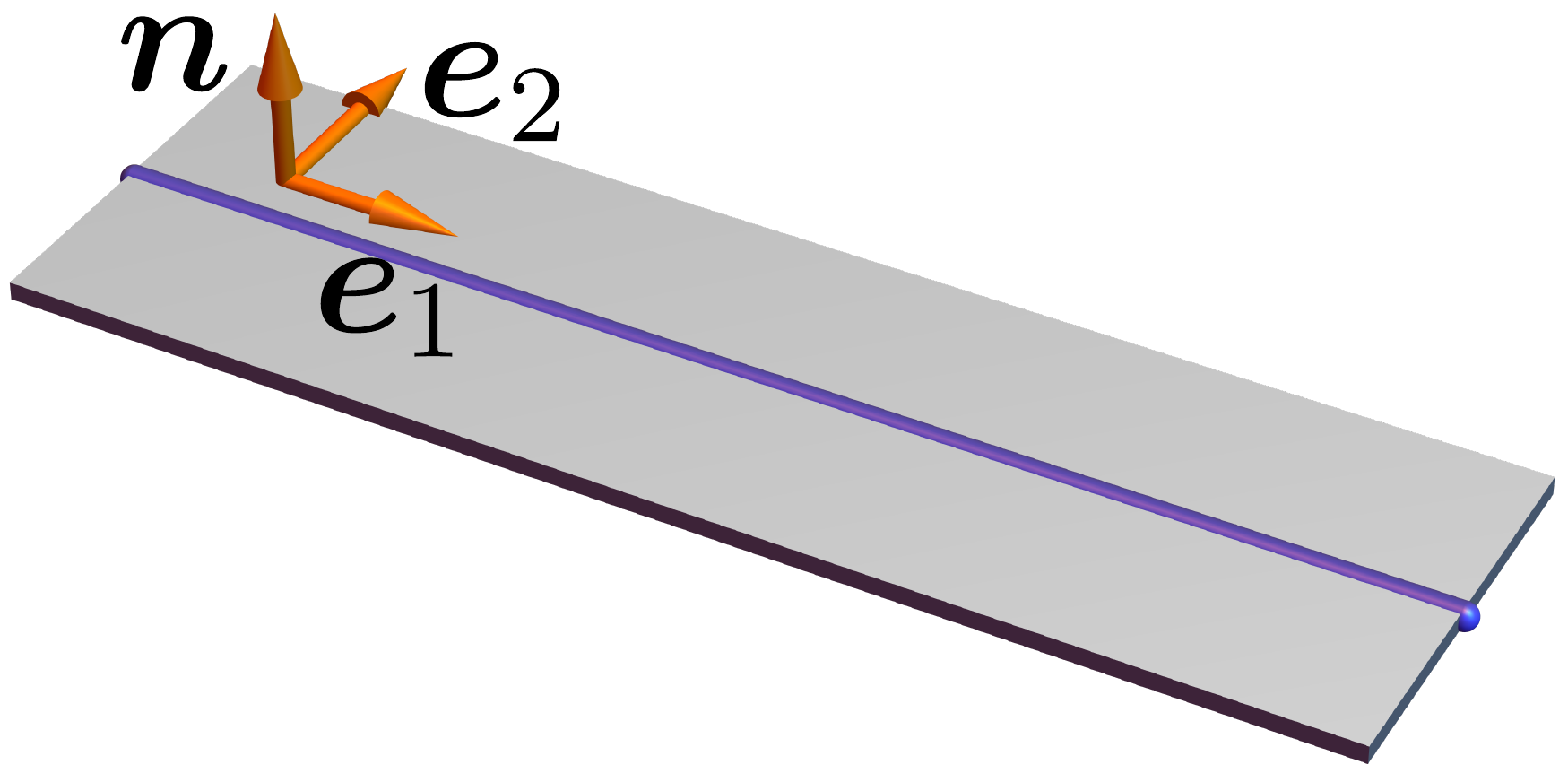} & \vspace{-2.5cm}\begin{itemize}
			\item $R_0\to\infty$;
			\item $P_0\to\infty$;
			\item $\dfrac{E_0}{hwL}=0$.
		\end{itemize} \\
		\hline \hline 
		\multirow{2}{*}[3cm]{\rotatebox{90}{\centering Interfacial DMI: $\mathcal{E}_\textsc{d}^\textsc{n}=m_n \vec{\nabla}\cdot\vec{m}-\vec{m}\cdot\vec{\nabla}m_n$}} & \centering $\vec{m}=\pm\vec{e}_1$ & \includegraphics[width=0.375\textwidth]{fig_straight.pdf} & \vspace{-2.5cm}\begin{itemize}
			\item $R_0\to\infty$;
			\item $P_0\to\infty$;
			\item $\dfrac{E_0}{hwL}=0$.
		\end{itemize} \\
		\cline{2-4}
		& \centering $\vec{m}=\pm\vec{n}$  & \includegraphics[width=0.375\textwidth]{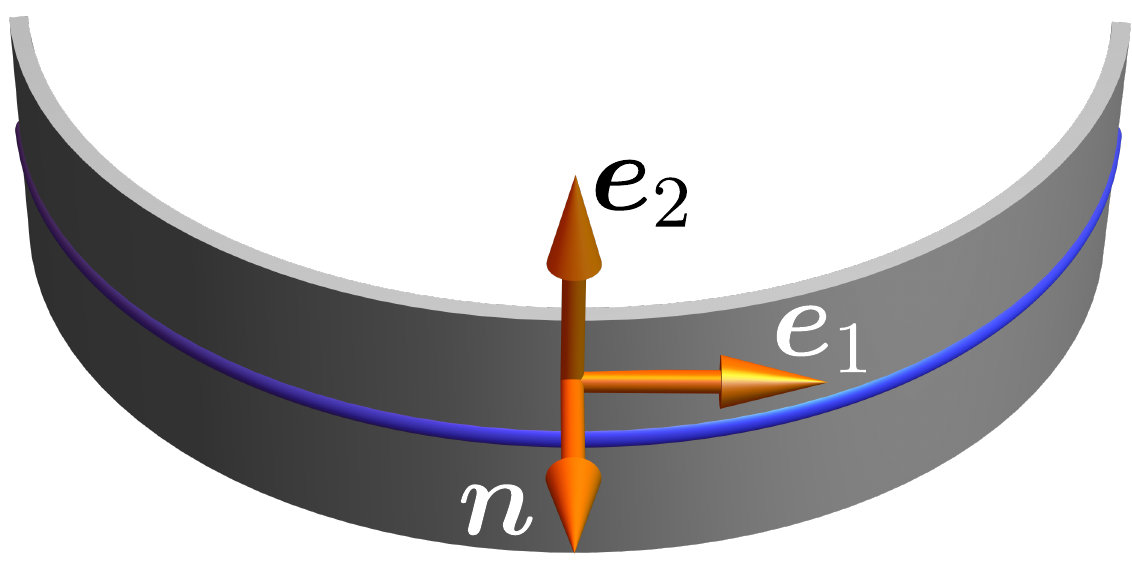} & \vspace{-2.5cm}\begin{itemize}
			\item $R_0^\textsc{cyl}=2\dfrac{A}{|D|}\left[1+\dfrac{Yh^2}{24A\left(1+\nu\right)}\right]$;
			\item $P_0^\textsc{cyl}=0$;
			\item $\dfrac{E_0^\textsc{cyl}}{hwL}=-\dfrac{D^2}{Y h^2}\dfrac{6\left(1+\nu\right)}{1+24A\left(1+\nu\right)/\left(Yh^2\right)}$.
		\end{itemize} \\
		\hline \hline
	\end{tabular} 
	\caption{\label{tab:states_suppl} Equilibrium states of the flexible ferromagnetic ribbon}
\end{table}

\subsection{Helicoid geometry}
\subsubsection{Case of easy-tangential anisotropy}\label{sec:hel_interfacial_tangential}
Here we will consider the ribbon with helicoid parametrization~\eqref{eq:helicoid_parametrization_suppl} and easy-tangential anisotropy. The energy~\eqref{eq:model_suppl} reads
\begin{equation}\label{eq:hel_energy_interfacial_suppl}
\begin{split}
\frac{E^\textsc{hel}}{hwL}=2A\frac{k}{w}\left[\frac{2kw\cos^2\theta}{\sqrt{4+\left(kw\right)^2}}+\text{arcsinh}\left(\frac{kw}{2}\right)\sin^2\theta\right]&\\
+K\left(1-\sin^2\theta\cos^2\phi\right)&+\frac{Y}{640\left(1-\nu^2\right)}\left(kw\right)^4+\frac{Yh^2}{6\left(1+\nu\right)}\frac{k}{w}\arctan\left(\frac{kw}{2}\right).
\end{split}
\end{equation}
Energy~\eqref{eq:hel_energy_interfacial_suppl} has zero DMI contribution, therefore one should expect not deformed ribbon. By minimizing~\eqref{eq:hel_energy_interfacial_suppl} we obtain that magnetic angles are defined as $\theta_0^\textsc{hel}=\pm\pi/2$ and $\cos\phi_0^\textsc{hel}=\mathfrak{C}$ with twist parameter $k=0$ which correspond to the straight and not deformed ribbon. Total energy of this state~\eqref{eq:hel_energy_interfacial_suppl} is zero $E^\textsc{hel}\left(\vec{e}_\textsc{a}=\vec{e}_1\right)=0$.

\subsubsection{Case of easy-normal anisotropy}\label{sec:hel_interfacial_normal}

Here we will use magnetization parametrization~\eqref{eq:magnetization_1_suppl} defined in Sec.~\ref{sec:dna_bulk_normal}. The energy for the helicoid geometry with easy-normal anisotropy can be written as
\begin{equation}\label{eq:hel_interfacial_norm_suppl}
\begin{split}
\frac{E^\textsc{hel}}{hwL}=2A\frac{k}{w}\left[\frac{2kw}{\sqrt{4+k^2w^2}}\sin^2\Theta\sin^2\Phi+\text{arcsinh}\frac{kw}{2}\left(\cos^2\Theta+\sin^2\Theta\cos^2\Phi\right)\right]\\
+K\left(1-\sin^2\Theta\sin^2\Phi\right)+\frac{Y}{640\left(1-\nu^2\right)}\left(kw\right)^4+\frac{Yh^2}{6\left(1+\nu\right)}\frac{k}{w}\arctan\left(\frac{kw}{2}\right).
\end{split}
\end{equation}
Similarly to the case described in~\ref{sec:hel_interfacial_tangential}, here we also have the deformation-independent zero DMI energy, therefore one should expect not deformed ribbon. By minimizing~\eqref{eq:hel_interfacial_norm_suppl} we obtain that magnetic angles are defined as $\Theta_0^\textsc{hel}=\pm\pi/2$ and $\Phi_0^\textsc{hel}=\pm\pi/2$ with twist parameter $k=0$ which correspond to the straight and not deformed ribbon.  Total energy of this state~\eqref{eq:hel_energy_interfacial_suppl} is zero $E^\textsc{hel}\left(\vec{e}_\textsc{a}=\vec{n}\right)=0$.

The results obtained in sections~\ref{sec:bulk_dmi} and \ref{sec:interfacial_dmi} presented in the Table~\ref{tab:states_suppl}.

\section{Details of numerical simulations}\label{sec:simulations}

In order to verify our analytical calculations we perform a set numerical simulations for a flexible ferromagnetic ribbon. We consider a ribbon with triangular lattice and lattice constant $a$. Each node has characterized by magnetic moment $\vec{m}_{\vec{q}}(t)$ which located at position $\vec{r}_{\vec{q}}(t)$. Here $\vec{q}=\left(i,j\right)$ is a two dimensional vector which defines magnetic moment and its position on a lattice with size $N_1\times N_2$ ($i\in\left[1,N_1\right]$ and $j\in\left[1,N_2\right]$). Magnetic moments are ferromagnetically coupled. The dynamics of magnetic subsystem is govern by Landau--Lifshitz equations
\begin{subequations}\label{eq:discrete_eq_suppl}
	\begin{equation}\label{eq:LL_suppl}
	\frac{\mathrm{d}\vec{m}_{\vec{q}}}{\mathrm{d}\tau}=\vec{m}_{\vec{q}}\times\frac{\partial\mathscr{H}}{\partial\vec{m}_{\vec{q}}}+\alpha\,\vec{m}_{\vec{q}}\times\left[\vec{m}_{\vec{q}}\times\frac{\partial\mathscr{H}}{\partial\vec{m}_{\vec{q}}}\right],
	\end{equation}
	while the dynamics of mechanical subsystem is described by the overdamped Newton equation
	\begin{equation}\label{eq:Newton_suppl}
	\eta\frac{\mathrm{d}\vec{r}_{\vec{q}}}{\mathrm{d}\tau}=-\frac{\partial\mathscr{H}}{\partial\vec{r}_{\vec{q}}},
	\end{equation}
\end{subequations}
where $\tau=\omega_0 t$ is a reduced time with $\omega_0 = |\gamma_0|K/M_s$, $\alpha$ and $\eta=\upsilon|\gamma_0|/M_s$ are dimensionless damping coefficients with $\upsilon$ being viscous damping parameter. $\mathscr{H}$ is a dimensionless energy normalized by $K$. We consider five contributions to the energy of the system
\begin{subequations}\label{eq:discrete_energy_suppl}
	\begin{equation}\label{eq:total_energy_suppl}
	\mathscr{H}=\mathscr{H}_\textsc{ex}+\mathscr{H}_\textsc{a}+\mathscr{H}_\textsc{d}+\mathscr{H}_\textsc{s}+\mathscr{H}_\textsc{b}.
	\end{equation}
	The first term in~\eqref{eq:total_energy_suppl} is the exchange energy
	\begin{equation}\label{eq:energy_exchange_suppl}
	\mathscr{H}_\textsc{ex}=-\frac{\ell^2}{a^2}\sum_{\vec{q}}\vec{m}_{\vec{q}}\cdot\vec{m}_{{\vec{q}}+{\vec{\delta}}},
	\end{equation}
	where $\vec{\delta}$ runs over nearest neighbors.
	
	The second term in~\eqref{eq:total_energy_suppl} is the anisotropy energy
	\begin{equation}\label{eq:energy_anisotropy_suppl}
	\mathscr{H}_\textsc{a}=-\sum_{\vec{q}}\left(\vec{m}_{\vec{q}}\cdot\vec{e}^\textsc{a}_{\vec{q}}\right)^2,
	\end{equation} 
	where $\vec{e}^\textsc{a}_{\vec{q}}$ is easy axis vector at node with coordinate $\vec{r}_{\vec{q}}$. For the case of easy tangential anisotropy $\vec{e}^\textsc{a}_{\vec{q}}=\left(\vec{r}_{\vec{q}+(i+1,j)}-\vec{r}_{\vec{q}}\right)/|(\vec{r}_{\vec{q}+(i+1,j)}-\vec{r}_{\vec{q}}|$, while for the easy-normal anisotropy $\vec{e}^\textsc{a}_{\vec{q}}=\vec{n}_{\vec{q}}=\sum_{k} \left[\mathcal{S}_k^{\triangle}(\vec{q})\vec{n}_k^{\triangle}(\vec{q})\right]/\left|\left[\mathcal{S}_k^{\triangle}(\vec{q})\vec{n}_k^{\triangle}(\vec{q})\right]\right|$, where $\vec{n}_k^{\triangle}(\vec{q})$ is $k$-th normal vector to the triangle with vertex in $\vec{r}_{\vec{q}}$ and $\mathcal{S}_k^{\triangle}(\vec{q})$ is an area of this triangle.
	
	The third term in Eq.~\eqref{eq:total_energy_suppl} is a  DMI energy
	\begin{equation}\label{eq:energy_dmi_suppl}
	\mathscr{H}_\textsc{d}=d_0\sum_{\vec{q}}\vec{d}_{\vec{q},\vec{\delta}}\cdot\left[\vec{m}_{\vec{q}}\times\vec{m}_{\vec{q}+\vec{\delta}}\right],
	\end{equation}
	with $d_0=D/K$, and $\vec{d}_{\vec{q},\vec{\delta}}$ being DMI vector. For the case of surface DMI $\vec{d}_{\vec{q},\vec{\delta}}=\vec{n}_{\vec{q}}\times \vec{u}_{\vec{q},\vec{\delta}}$ with $\vec{u}_{\vec{q},\vec{\delta}} = \left(\vec{r}_{\vec{q}+\vec{\delta}}-\vec{r}_{\vec{q}}\right)/\left|\vec{r}_{\vec{q}+\vec{\delta}}-\vec{r}_{\vec{q}}\right|$ being a unit vector which connects two nearest neighbors. For other DMI symmetry $\vec{d}_{\vec{q},\vec{\delta}}=\vec{u}_{\vec{q},\vec{\delta}}$.
	
	The fourth term in Eq.~\eqref{eq:total_energy_suppl} is a  stretching energy
	\begin{equation}\label{eq:energy_stretching_suppl}
	\mathscr{H}_\textsc{s}=\lambda\sum_{\vec{q}}\left[\left|\vec{r}_{\vec{q}+\vec{\delta}}-\vec{r}_{\vec{q}}\right|-a\right]^2,
	\end{equation}
	with $\lambda=\Lambda/K$ being a stretching parameter.
	
	The last term in the Eq~\eqref{eq:total_energy_suppl} is a bending energy
	\begin{equation}\label{eq:energy_bendingg_suppl}
	\mathscr{H}_\textsc{b}=-\beta\sum_{p}\vec{n}_{p}^\triangle\cdot\vec{n}_{p+k}^\triangle,
	\end{equation}
	where we summarize over all triangle in a lattice and $k$ defines nearest triangles~(each triangle has three nearest triangles except ones which are located near the edges). Here $\beta=B/K$ defines the bending constant.
\end{subequations}

\subsection{Matching of elastic parameters in discrete and continuum models}\label{sec:mathc}

\begin{figure}
	\includegraphics[width=\textwidth]{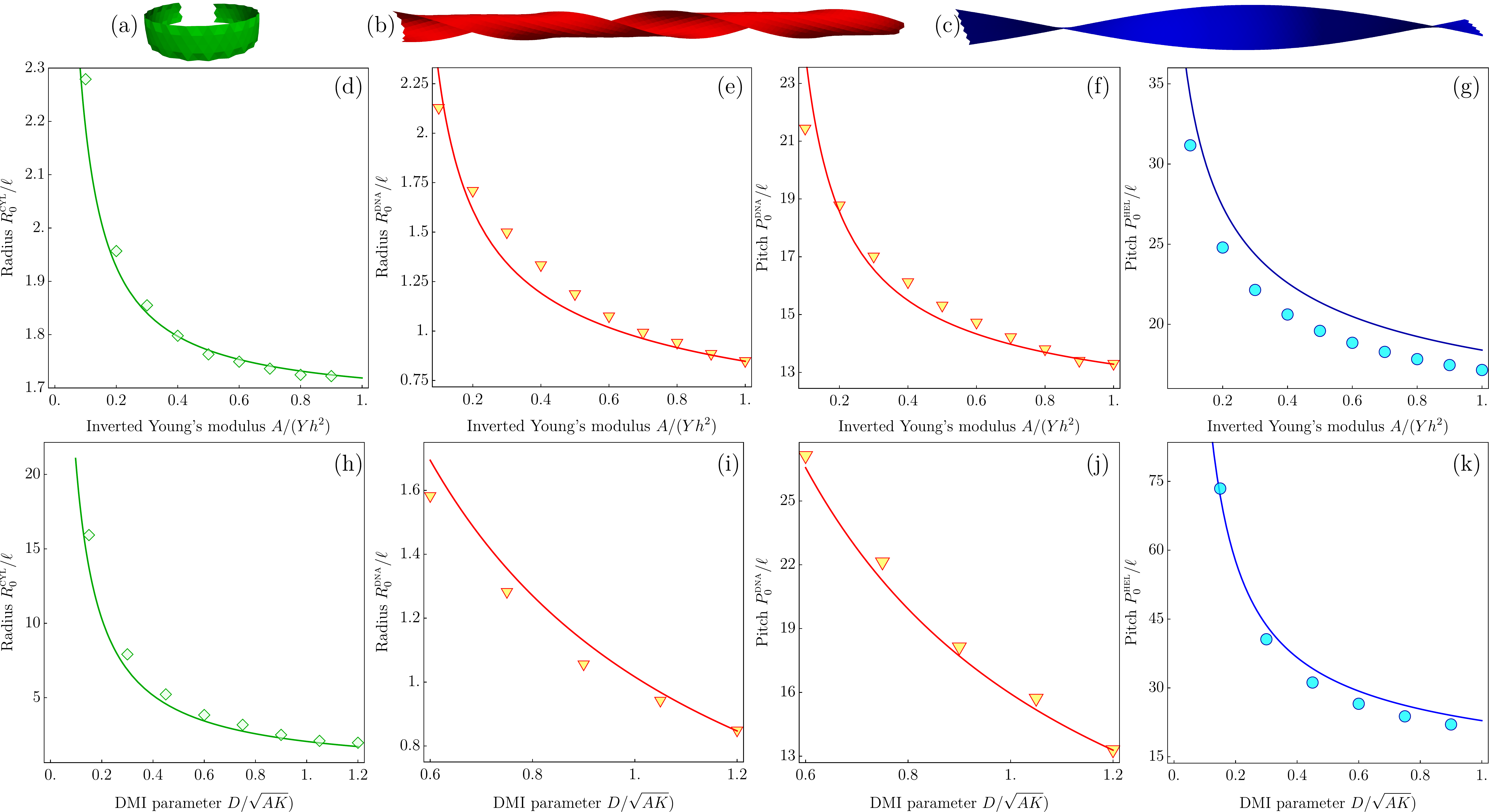}
	\caption{\label{fig:parameters_suppl}%
		(Color online) \textbf{Geometrical parameters of flexible ferromagnetic ribbons}: (a) -- (c) are shapes of flexible ferromagnetic ribbons obtained by means of numerical simulations. (d)-(g) Geometrical parameters as functions of Young's modulus: (d) -- cylindrical state; (e) and (f) DNA-like state; (g) helicoid state. (h)-(k) Geometrical parameters as functions of DMI constant: (d) -- cylindrical state; (e) and (f) DNA-like state; (g) helicoid state. In all cases $\nu=1/3$, $\ell=4a$.}
\end{figure} 

In order to find correspondence between the elastic parameters~$Y$ and $\nu$, effective thickness $h$ in continuum model~\eqref{eq:model_suppl} and parameters $\Lambda$ and $B$ in discrete model~\eqref{eq:discrete_energy_suppl} we performed two types of numerical simulations of nonmagnetic flexible ribbon. Firstly, we simulate the cylindrical surface, which is free of stretching energy~($\overline{g}_{\alpha\beta}=g_{\alpha\beta}$), and compared the energies given by Eq.~\eqref{eq:bending_suppl} and energy obtained from simulations. The second type of simulations was performed for a flat and straight ribbon. We applied external force to the edges of the ribbon in opposite directions in order to stretch it. In this case the bending energy of the ribbon is equal to zero. By comparing initial and final geometrical parameter of the ribbon we calculated the Poisson ration for a triangular lattice $\nu^\triangle=-\frac{w'-w_0}{w_0}\frac{L_0}{L'-L_0}$, where $w_0$ and $L_0$ are geometrical parameter before deformation, while  $w'$ and $L'$ are geometrical parameters after deformation. For a triangular lattice  we obtained the Poisson ratio equal to $\nu^\triangle=1/3$. Components of actual metric $g_{\alpha\beta}$ were calculated as $g_{11}=L'/L_0>1$, $g_{22}=w'/w_0<1$, and $g_{12}=g_{21}=0$. Results of simulations are presented in Table~\ref{tab:results}.

\begin{table}[t]
	\begin{tabular}{|p{0.025\columnwidth}|p{0.1\columnwidth}|p{0.1\columnwidth}|p{0.1\columnwidth}|}
		\hline
		\centering $N_2$& \centering $\nu^\triangle$ & \centering $\Lambda/\left(Yh\right)$ &  $B/\left(Yh^3\right)$ \\
		\hline
		\centering 3& \centering 0.336974 & \centering 0.294375 & 0.0585751 \\
		\hline
		\centering 5& \centering 0.335962 & \centering 0.353973 & 0.0561083 \\
		\hline
		\centering 7& \centering 0.334573 & \centering 0.379484 & 0.0554347 \\
		\hline
		\centering 9& \centering 0.333438 & \centering 0.385654 & 0.0541765 \\
		\hline
		\centering 11& \centering 0.333345 & \centering 0.356434 & 0.0540165 \\
		\hline 
	\end{tabular} 
	\caption{\label{tab:results}Matching between continuum and discreet models.}
\end{table}

\subsection{Simulation of relaxation dynamics}\label{sec:relax}

In order to verify the equilibrium state of the ribbon we performed a set of simulations for various range of mechanical, magnetic, and geometrical parameters. In each simulation we start with a geometry of straight and flat ribbon. The initial magnetization state was defined by the easy-axis anisotropy direction, i.e. $\vec{m}_{\vec{q}}\left(t=0\right)=\vec{e}_{\vec{q}}^\textsc{a}$. The simulations were performed on a long time interval with $\alpha = 0.5$ and $\upsilon=0.01 M_s/|\gamma_0|$. In all cases we consider $\ell=4a$. Geometrical parameters as functions of Young's modulus and DMI constant are presented in Fig.~\ref{fig:parameters_suppl}.

For the case of bulk DMI~\eqref{eq:dmi_bulk_suppl} we performed additional simulations in order to build phase diagrams. We use magnetization distribution and positions of nodes from DNA-like state and use it as initial conditions for cases where our analytical calculations predict the helicoid state as equilibrium. The final state with lowest energy is considered as equilibrium state, see Fig.~\ref{fig:states}.

\subsection{Simulation with external magnetic field}\label{sec:field}

We also studied the influence of external magnetic field on the equilibrium states studied above. The magnetic field was applied along the $\hat{\vec{z}}$ axis, i.e. $\vec{H}=H\hat{\vec{z}}$, for the DNA-like and helicoid states, while for the cylindrical state $\vec{H}=H\hat{\vec{x}}$. The interaction with magnetic field is represented by the Zeeman term $\mathscr{H}_\textsc{z}=-1/\left(2K/M_s\right)\sum_{\vec{q}}\vec{\vec{m}_\vec{q}\cdot\vec{H}}$, where $H$ is the magnetic field amplitude. Results of numerical simulations are presented in Fig.~\ref{fig:field}.

\subsection{Movies of the DMI and field induced deformations}\label{sec:movis}

\titleformat{\subsubsection}[runin]
	{\normalfont\normalsize\bfseries}{\thesubsubsection}{1.em}{}
For better illustration of DMI and field induced deformations we provide four movies constructed from our numerical simulations:

\addcontentsline{toc}{subsubsection}{Appendices}
\renewcommand{\thesubsubsection}{Movie A}
\subsubsection{}\label{sec:video_bloch_free} Video file "\texttt{movie\_bloch\_dmi\_free.mp4}" shows the spontaneous deformation of two ferromagnetic ribbons with DMI of Bloch type and easy-tangential anisotropy. We start with straight ribbons placed in $\vec{z}\vec{y}$ plane magnetized allong $\vec{z}$ direction (tangential direction). The final state of the ribbons is defined by the DMI strength, i.e. for $D=1.2\sqrt{AK}$ we have DNA-like state, while for $D=0.45\sqrt{AK}$ we have helicoid state. In movie we have ribbons with the following parameters $L=25\ell$, $w\approx2\ell$, $h=0.05\ell$, and $A/\left(Yh^2\right)=1$.

\renewcommand{\thesubsubsection}{Movie B}
\subsubsection{}\label{sec:video_bloch_field} Video file "\texttt{movie\_bloch\_dmi\_field.mp4}" illustrates the field induced deformation of ribbons obtained from numerical simulations presented in movie \texttt{movie\_bloch\_dmi\_free.mp4}. Here magnetic field is applied along the $\vec{z}$ direction with amplitude $H/\left(2K/M_s\right) = 0.5$.

\renewcommand{\thesubsubsection}{Movie C}
\subsubsection{}\label{sec:video_neel_free} Video file "\texttt{movie\_neel\_dmi\_free.mp4}" shows the spontaneous deformation of the ferromagnetic ribbons with DMI of Ne{\'e}l type and easy-normal anisotropy. Here, we start with straight ribbons placed in $\vec{x}\vec{y}$ plane magnetized allong $\vec{x}$ direction (normal direction).In movie we have ribbon with the following parameters $L=10\ell$, $w\approx2\ell$, $h=0.05\ell$, $D=1.2\sqrt{AK}$, and $A/\left(Yh^2\right)=1$.

\renewcommand{\thesubsubsection}{Movie D}
\subsubsection{}\label{sec:video_neel_field} Video file "\texttt{movie\_neel\_dmi\_field.mp4}" illustrates the field induced deformation of ribbons obtained from numerical simulations presented in movie \texttt{movie\_neel\_dmi\_free.mp4}. Here magnetic field is applied along the $\vec{x}$ direction with amplitude $H/\left(2K/M_s\right) = 0.5$.

\end{widetext}

\end{document}